\documentclass[12pt,authoryear]{article}

\usepackage{amsmath,amsfonts,amssymb}
\usepackage{lineno}
\usepackage{pgf}
\usepackage{tikz}
\usepackage{subfigure}
\usepackage{algorithm}
\usepackage{algorithmic}
\usepackage[colorlinks=true]{hyperref}
\usepackage{color}
\begin{document}

%\begin{frontmatter}


% \ead{eliberge@univ-lr.fr}

\title{Fluid Structure Interaction modelling using Lattice Boltzmann and Volume Penalization method}


\author{Erwan Liberge, Claudine B\'eghein \\
LaSIE UMR 7356 CNRS, Universit\'e de La Rochelle,\\  Avenue Michel Cr\'epeau, 17042 La Rochelle Cedex \\ France}
\date{eliberge@univ-lr.fr}
%\address{LaSIE UMR 7356 CNRS, Universit\'e de La Rochelle,\\  Avenue Michel Cr\'epeau, 17042 La Rochelle Cedex \\ France}
\maketitle
\begin{abstract}
This paper proposes an approach combining the Volume Penalization (VP) and the the Lattice Boltzmann method (LBM) to compute fluid structure interaction involving rigid bodies. The method consists in adding a force term in the LBM formulation, and thus considers the rigid body similar to particular porous media. Using a characteristic function for the solid domain avoids the expensive tracking of the fluid-solid interface employed commonly in LBM to treat FSI problems. The method is applied to three FSI problems and solved using a Graphics Processor Units (GPU) device. The applications focus on the capacity of the method to compute the drag and lift coefficients for various cases : the imposed displacement of a cylinder, the particle sedimentation at a very low Reynolds number and the VIV of a cylinder in a transverse fluid flow. For all cases the VP-LBM approach yields results which are in a good agreement with those of literature. 
\end{abstract}

%\begin{keyword}
\begin{paragraph}{Keyword}
Lattice Boltzmann Method, Fluid Structure Interaction, Volume Penalization, Vortex Induced Vibration (VIV)
\end{paragraph}

%\end{keyword}

%\end{frontmatter}

%\linenumbers

\section{Introduction}
\label{sec:1}

In the following section, the mathematical background is presented. This part deals with the Lattice Boltzmann Method and more particularly the Multiple Relaxation Time (MRT) approach, the Volume Penalization and how these two methods are combined. The last section presents computational applications computed on a GPU device. For the first tested case, the imposed displacement of a cylinder in a transverse fluid flow at a Reynolds of $100$, is not a real case of FSI. This application enables to validate the capacity of the method to compute drag and lift forces. The second example deals with particle sedimentation under gravity, the particle is free to move under fluid and gravity constraints and complex trajectories can be obtained. The last application focuses on the behavior of the VP-LBM method for FSI problems when a parameter is varied. The stiffness parameter of a free oscillating cylinder in a transverse fluid flow is changed, and the capacity of the VP-LBM to reproduce results from the literature is evaluated.

\section{Governing equations}
\label{sec:2}
In this section, the numerical models are exposed. The following notations are used : $\rho$ and $\mathbf{u}$ are the macroscopic density and velocity, and bold characters denote vectors.
\subsection{Volume penalization}
Let us consider a fluid domain $\Omega_f$, a solid domain $\Omega_s$, $\Gamma$ the fluid-solid interface, and let us note $\Omega = \Omega_f \cup \Omega_s \cup \Gamma$. The Volume Penalization (VP) method consists in extending the Navier-Stokes equations on the whole domain $\Omega$, and considering the solid domain as a porous medium with a very small permeability. The method was introduced by Angot et al. \cite{Angot1999497} and already applied to macroscopic equations for moving bodies  \cite{Kadoch20124365}. The small permeability of the solid domain is  modelled using a penalization coefficient, hence the desired boundary conditions at the fluid-solid interface are naturally imposed. With this method, the incompressible Navier-Stokes equations are written as follows :
\begin{equation}
\begin{array}{ll}
\displaystyle \nabla \cdot \mathbf{u}=0 \\
\displaystyle \frac{\partial \mathbf{u}}{\partial t}+\mathbf{u}\cdot \nabla \mathbf{u} = \displaystyle- \frac{1}{\rho}\nabla p + \nu \nabla^2 \mathbf{u} -\dfrac{\chi_{\Omega_s}}{\eta} \left(\mathbf{u} - \mathbf{u_s}\right)  \\
\end{array}
\end{equation} 
where  
\begin{equation}
\chi_{\Omega_s}\left(\boldsymbol{x},t\right)=\left\lbrace \begin{array}{ll} 1 \;\; \textrm{if} \;\; \boldsymbol{x} \in \Omega_s\left( t \right) \\
0 \;\; \textrm{otherwise}
\end{array}
 \right. ;  \hspace{1cm}\eta \ll 1\;\; \textrm{penalization factor}
 \label{eq:fcar}\end{equation} 
$\mathbf{u}$ denotes the velocity field, $p$ is the pressure field, $\rho$ and $\nu$ are the density and the viscosity of the fluid. $\mathbf{F}=\dfrac{\chi_{\Omega_s}}{\eta} \left(\mathbf{u} - \mathbf{u_s}\right)  $ is the penalization term, and $ \mathbf{u_s}$ is the velocity field in the solid domain. 
\subsection{Lattice Boltzmann method}
Based on the Boltzmann equation (equation (\ref{eqB}))  proposed in the context of the Kinetic Gaz Theory by L. Boltzmann in 1870, the Lattice Boltzmann Method has been successfully used  to model fluid flow since the 90's 
\begin{equation}
\label{eqB}
\displaystyle \dfrac{\partial f }{\partial t } + \mathbf{c}\cdot \nabla_x f= \Omega\left(f \right) 
\end{equation}
This equation models the transport of  $f\left(\boldsymbol{x},t,\mathbf{c}\right)$, a probability density function to find a particle at location $\boldsymbol{x}$ and time $t$ with the velocity $\mathbf{c}$, $\Omega\left(f \right) $ being a collision operator. The link between the Boltzmann equation and the Navier-Stokes equations is well-known since the Chapmann-Enskog expansion proposed in 1915. \\ 
The Lattice Boltzmann method considers the discretization of equation (\ref{eqB}) according to space and velocity and leads to the following discretized equations :
\begin{equation}
\label{eqBd}
\displaystyle f_{\alpha} \left(\boldsymbol{x}+\mathbf{c_\alpha}\triangle t, t +\triangle t \right) -f_{\alpha} \left(\boldsymbol{x}, t  \right) = \Omega_\alpha \left(f \right) +\triangle t F_\alpha 
\end{equation}
where $ \displaystyle  f_{\alpha} \left(\boldsymbol{x}, t  \right) = \displaystyle f \left(\boldsymbol{x},\mathbf{c_{\alpha}}, t  \right) $, $ \displaystyle F_\alpha $ is a forcing term related to the discrete velocity $\mathbf{c_\alpha}$. 
 The first model proposed by Bhatnagar et al. \cite{Bhatnagar1954511} is the BGK model which is based on a linear collision operator with a single relaxation time :
\begin{equation}
\Omega_\alpha \left(f\right)= -\dfrac{1}{\tau} \left(f_\alpha \left( \boldsymbol{x},t\right) -f ^{\textrm{eq}}_\alpha \left( \boldsymbol{x},t\right) \right)
\end{equation}
where $f ^{\textrm{eq}}$ is the equilibrium function and $\tau$ is the non dimensional relaxation time which is linked to the fluid viscosity as follows (\ref{tauvisco}).
\begin{equation}\label{tauvisco}
\nu =c^2_s\triangle t \left( \tau -\dfrac{1}{2}\right)
\end{equation} 
We propose in this paper to use the Multiple Relaxation Time (MRT) model, introduced by d'Humi\`eres \cite{dh92} for stability reasons. This scheme consists in using a transformation matrix $M$ to work with macroscopic quantities in the moment space. For the MRT model, the collision operator is :
\begin{equation}
\Omega_\alpha \left(f\right)= -\sum_\beta \left( M^{-1} S M \right)_{\alpha \beta} \left(f_\beta \left( \boldsymbol{x},t\right) -f ^{\textrm{eq}}_\beta \left( \boldsymbol{x},t\right) \right)
\end{equation}
where $S$ is a diagonal relaxation matrix.

The transformation matrix $M$ enables to express the moments $m_{\alpha}(\boldsymbol{x},t)$ according 
to the distribution functions $f_{\alpha}(\boldsymbol{x},t)$. The lattice Boltzmann equation with a source term, as given by Lu et al. \cite{Lu2012}, becomes:
\begin{multline}
 |f_{\alpha}(\boldsymbol{x}+\mathbf{c}_{\alpha} \Delta t,t+ \Delta t)\rangle-|f_{\alpha}(\boldsymbol{x},t)\rangle=\\
  \displaystyle -M^{-1} \left( S \left( |m_{\alpha}\left( \boldsymbol{x},t \right)\rangle-|m_{\alpha}^{eq}\left(\boldsymbol{x},t\right)\rangle\right) -\left(I-\dfrac{S}{2}\right)M\Delta t |F_{\alpha}\left(\boldsymbol{x},t\right)\rangle \right)
 \label{equa_lbl_mrt}
\end{multline}
In this equation, $|\bullet \rangle$ denotes a column vector and $I$ is the identity matrix. The moments $|m_{\alpha}\left(\boldsymbol{x},t\right)\rangle=\left(m_0,m_1,\ldots,m_8\right)^T$ are deduced from:
\begin{equation}
 |m_{\alpha}(\boldsymbol{x},t)\rangle= M|f_{\alpha}(\boldsymbol{x},t)\rangle \Rightarrow |f_{\alpha}(\boldsymbol{x},t)\rangle=M^{-1}|m_{\alpha}(\boldsymbol{x},t)\rangle
\end{equation}

The model used most commonly to simulate two-dimensional flows is the nine-velocity square lattice model D2Q9 (figure \ref{d2q9}) \cite{Yu2003329}. 
\begin{figure}[!htbp]
\center 
\begin{tikzpicture}
%\pgftext[at=\pgfpoint{1cm}{-1cm},left,base]{\pgfuseimage{D2Q9}}
\draw [latex-latex] (0.05,2.05) -- (0.95,2.05);
\node [above] at (0.5,2) { $\triangle x$};
\draw [latex-latex] (0.05,1.05) -- (0.05,1.95);
\node   at (-0.5,1.5) { $\triangle y$};
%\node [above right] at (2,-2) {D2Q9};
\node  [blue] at (2.6,1.5) {$\mathbf{c_0}$};
\node  [blue,below right] at (4,1) {$\mathbf{c_1}$};
\node  [blue,above right] at (3,2) { $\mathbf{c_2}$};
\node  [blue,below left] at (2,1) { $\mathbf{c_3}$};
\node  [blue,below right] at (3,0) { $\mathbf{c_4}$};
\node  [blue,above right] at (4,2) {$\mathbf{c_5}$};
\node  [blue,above left] at (2,2) { $\mathbf{c_6}$};
\node  [blue,below left] at (2,0) { $\mathbf{c_7}$};
\node  [blue,below right] at (4,0) { $\mathbf{c_8}$};
\draw (-0.5,-0.5) grid (5.5,3.2);
\draw [blue,thick,-latex] (3,1) -- (4,1);
\draw [blue,thick,-latex] (3,1) -- (4,2);
\draw [blue,thick,-latex] (3,1) -- (3,2);
\draw [blue,thick,-latex] (3,1) -- (2,2);
\draw [blue,thick,-latex] (3,1) -- (2,1);
\draw [blue,thick,-latex] (3,1) -- (2,0);
\draw [blue,thick,-latex] (3,1) -- (3,0);
\draw [blue,thick,-latex] (3,1) -- (4,0);
\end{tikzpicture}
\caption{\label{d2q9} Discrete velocities of the D2Q9 model}
\end{figure}
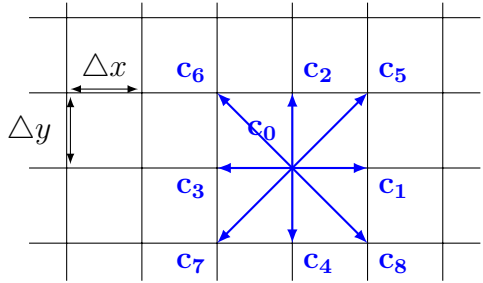
\begin{equation}
\mathbf{c_\alpha} = \left\lbrace
\begin{array}{lll}
\left(0,0 \right) & \alpha =0\\
\left( \textrm{cos} \left(   \left(  \alpha -1 \right)  \dfrac{\pi}{2} \right) ,  \textrm{sin} \left(   \left(  \alpha -1 \right)  \dfrac{\pi}{2} \right) \right)c & \alpha =1,2,3,4 \\
\left( \textrm{cos} \left(   \left( 2 \alpha -9 \right)  \dfrac{\pi}{4} \right) ,  \textrm{sin} \left(   \left(2  \alpha -9 \right)  \dfrac{\pi}{4} \right) \right)\sqrt{2}c& \alpha =5,6,7,8
\end{array}
 \right.
\end{equation}
Where $c=\dfrac{\triangle x }{\triangle t}$. Usually $\triangle x=\triangle y= \triangle t=1$ are chosen. In addition, the equilibrium distribution function for the D2Q9 model is :
\begin{equation}
\displaystyle f^{\textrm{eq}}_\alpha  = \displaystyle \omega_\alpha \rho \left( 1 + \dfrac{\mathbf{c_\alpha} \cdot \mathbf{u}}{c^2_s} +  \dfrac{ \mathbf{u}  \mathbf{u}  :  \left( \mathbf{c_\alpha} \mathbf{c_\alpha} - c^2_s I \right) }{2c^4_s}  \right),
\end{equation}
and the forcing term $ \displaystyle F_\alpha $ in equation (\ref{eqBd}) is \cite{guo_discrete_2002} :
\begin{equation}
\displaystyle F_\alpha = \displaystyle \left( 1 - \dfrac{1}{2 \tau }\right) \omega_\alpha \left( \dfrac{\mathbf{c_\alpha} - \mathbf{u}}{c^2_s}  +  \dfrac{\mathbf{c_\alpha} \cdot  \mathbf{u}}{c^4_s} \mathbf{c_\alpha} \right) \cdot \rho \mathbf{F}  
\end{equation}
where $\omega_\alpha$ are the weighting coefficients commonly used for the D2Q9 model :
\begin{equation}
\omega_\alpha = \left\lbrace \begin{array}{lll}
4/9, & & \alpha = 0\\
1/9, &&  \alpha = 1,2,3,4\\
1/36& & \alpha = 5,6,7,8\\
\end{array} \right.,
\end{equation}
$c_s$ is the speed of sound, which for the D2Q9 model is $c_s=\dfrac{c}{ \sqrt{3}}$.

For the D2Q9 model the corresponding equilibria in the moment space are given by :
\begin{equation}
\begin{array}{ll}
 |m_{\alpha}^{eq}\rangle &= \displaystyle  \left(\rho, e^{eq}, \varepsilon^{eq},j^{eq}_x,q^{eq}_x,j^{eq}_y,q^{eq}_y,p^{eq}_{xx},p^{eq}_{xy} \right)\\
& = \displaystyle \rho \left(1,-2+3(u^2_x+u^2_y),1-3(u^2_x+u^2_y),u_x,-u_x,u_y,-u_y,u^2_x-u^2_y,u_xu_y \right)^T
\end{array} 
\end{equation}
where $u_x$ (respectively $u_y$) is the horizontal (respectively vertical) component of velocity $\mathbf{u}$.

The transformation matrix $M$ is :
\begin{equation}
\begin{split}
\begin{array}{cccccccccccccc}
 \displaystyle M=
\begin{pmatrix}
1 & \hspace{0.2cm} 1 \hspace{0.2cm} & 1 \hspace{0.2cm} & 1 \hspace{0.2cm} & \hspace{0.2cm} 1 & \hspace{0.2cm} 1 & \hspace{0.2cm} 1 & \hspace{0.2cm} 1 &\hspace{0.2cm} 1   \\
-4 & -1 & -1 & -1 & -1 & 2 & 2 & 2 & 2 \\
4 & -2 & -2 & -2 & -2 & 1 & 1 & 1 & 1 \\
0 & 1 & 0 & -1 & 0 & 1 & -1 & -1 & 1 \\
0 & -2 & 0 & 2 & 0 & 1 & -1 & -1 & 1 \\
0 & 0 & 1 & 0 & -1 & 1 & 1 & -1 & -1 \\
0 & 0 & -2 & 0 & 2 & 1 & 1 & -1 & -1 \\
0 & 1 & -1 & 1 & -1 & 0 & 0 & 0 & 0 \\
0 & 0 & 0 & 0 & 0 & 1 & -1 & 1 & -1 
\end{pmatrix}
 \end{array},
\end{split}
\label{eq1.7}
\end{equation}
and $S$ is a diagonal
 matrix that contains the relaxation rates of each moment:\\
\begin{equation}
 S=diag (s_0,s_1,s_2,s_3,s_4,s_5,s_6,s_7,s_8)  %=diag(0,s_n,s_n,0,s_q,0,s_q,s_n,s_n)
\end{equation}

At each time-step, the LBM algorithm consists in solving first a collision step : 
\begin{equation}
   | \tilde{f}_{\alpha} \rangle=   \displaystyle M^{-1} \left( |m_{\alpha} \rangle - S \left( |m_{\alpha}\rangle-|m_{\alpha}^{eq}\rangle\right) -\left(I-\dfrac{S}{2}\right)M\Delta t |F_{\alpha} \rangle \right)
  \label{eq:collision}
\end{equation}
%
%\begin{equation}
%  \tilde{f}_{\alpha}\left( \boldsymbol{x},t\right)=   \displaystyle M^{-1} \left( |m_{\alpha}\left( \boldsymbol{x},t \right)\rangle - S \left( |m_{\alpha}\left( %\boldsymbol{x},t \right)\rangle-|m_{\alpha}^{eq}\left(\boldsymbol{x},t\right)\rangle\right) -\left(I-S/2\right)M\Delta t |F_{\alpha}\left(\boldsymbol{x},t\right)\rangle \right)
%  \label{eq:collision}
%\end{equation}
%\begin{equation}
%  \tilde{f}_{\alpha}\left( \boldsymbol{x},t\right)=f_\alpha \left( \boldsymbol{x},t\right) - \dfrac{1}{\tau} \left(f_\alpha \left( \boldsymbol{x},t\right) -f ^{\textrm{eq}}_\alpha \left( \boldsymbol{x},t\right) \right)+\triangle t F_\alpha
%  \label{eq:collision}
%\end{equation}
and next a streaming step :
\begin{equation}
 | \displaystyle f_{\alpha} \left(\boldsymbol{x}+\mathbf{c_\alpha}\triangle t, t +\triangle t \right)  \rangle = | \tilde{f}_{\alpha} \left(\boldsymbol{x}, t  \right)  \rangle
  \label{eq:streaming}
\end{equation}
Finally, the macroscopic quantities are computed according to the following expressions :
\begin{equation}\label{updatestep}
\rho=\sum_\alpha f_\alpha \hspace{1cm} \rho \mathbf{u}=\sum_\alpha \mathbf{c_\alpha} f_\alpha +\dfrac{\triangle t}{2} \rho \mathbf{F}  
\end{equation}

In the present approach, the volume penalization term is added :
\begin{equation}\label{updatestep2}
 \rho \mathbf{u}=\sum_\alpha \mathbf{c_\alpha} f_\alpha -\dfrac{\triangle t}{2} \rho \dfrac{\chi_{\Omega_s}}{\eta} \left(\mathbf{u} - \mathbf{u_s}\right)  
\end{equation}
To avoid instabilities, the term including $\mathbf{u}$ in the penalization force is moved to the left hand side of equation (\ref{updatestep2}) 
\begin{equation}\label{updatestep3}
 \rho \left(1 + \dfrac{\triangle t}{2} \dfrac{\chi_{\Omega_s}}{\eta}  \right) \mathbf{u}=\sum_\alpha \mathbf{c_\alpha} f_\alpha +\dfrac{\triangle t}{2} \rho \dfrac{\chi_{\Omega_s}}{\eta}  \mathbf{u_s}  
\end{equation}

 This leads to the modified update step to compute the macroscopic velocity field :
\begin{equation}\label{updatestepm}
 \displaystyle \mathbf{u}= \displaystyle  \dfrac{ \displaystyle \sum_\alpha \mathbf{c_\alpha} f_\alpha   +   \dfrac{\triangle t}{2}  \dfrac{\chi_{\Omega_s}}{\eta}  \rho \mathbf{u_s}  }{\rho +    \dfrac{\triangle t}{2}  \dfrac{\chi_{\Omega_s}}{\eta}  \rho } 
\end{equation}
In the fluid domain, where $ \chi_{\Omega_s} = 0$ the classical LBM equation is obtained whereas in the solid domain, where $ \chi_{\Omega_s} = 1$, equation (\ref{updatestepm}) forces the velocity field to approach $ \mathbf{u_s}$.

\subsection{Structure displacement}
In this study, only structures which can be modelled as rigid bodies coupled with springs and dampers are considered. The center of gravity moves according to the following equation :
\begin{equation}\label{xstru}
 m \dfrac{d^2 \boldsymbol{x_G}}{d t^2} + c \dfrac{d  \boldsymbol{x_G}  }{d t} + k \left( \boldsymbol{x_G} -  \boldsymbol{x_0} \right) = \boldsymbol{\mathcal{F}_f} +\boldsymbol{\mathcal{F}_{ext}}
\end{equation}
where $ \boldsymbol{x_G}$ are the coordinates of the center of gravity of the solid body, $ \boldsymbol{x_0}$ is the equilibrium position, $m$ is the mass, $c$ and $k$ are the damping and stiffness coefficients,  $\boldsymbol{\mathcal{F}_f}$ are the fluid forces and $\boldsymbol{\mathcal{F}_{ext}}$ are external forces (gravity for example).
The rotation of the solid is solved using the following equation : 
\begin{equation}\label{thetastru}
I \dfrac{d^2 \boldsymbol{\theta}}{d t^2} + c_\theta \dfrac{d \boldsymbol{\theta}}{dt} + k_\theta \left(\boldsymbol{\theta} - \boldsymbol{\theta_0} \right) = \boldsymbol{\mathcal{T}_f}+\boldsymbol{\mathcal{T}_{ext}}
\end{equation}
where $\boldsymbol{\theta}$ is the rotation vector, $I$ is the inertia moment of the body, $c_\theta$ and $k_\theta$ are the damping and stiffness coefficients for rotation, $\boldsymbol{\mathcal{T}_f}$ is the torque induced by the fluid and $\boldsymbol{\mathcal{T}_{ext}}$ is an external torque.

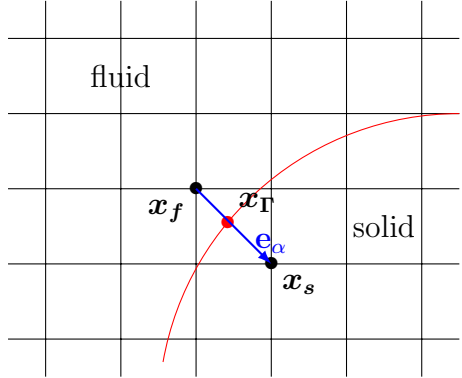
\begin{figure}[!htpb]
  \centering
 \begin{tikzpicture}
  \draw (-0.5,-0.5) grid (5.5,4.5);
  %\draw (0,3) -- (10,3)
  \draw  [red] (5.5,3) arc [radius=4, start angle=90, end angle= 170];
  \node [below right] at (3,1) {$\boldsymbol{x_s}$};
  \node at (3,1) {$\bullet$};
\node [below left] at (2,2) {$\boldsymbol{x_f}$};
   \node at (2,2) {$\bullet$};
   \node [above right] at (2.42,1.55) {$\boldsymbol{\displaystyle x_\Gamma}$};
     %\node [thick] at (2.42,1.55) {$\times$};
      \node [red] at (2.42,1.55) {$\bullet$};
  \draw [blue,thick,-latex] (2,2) -- (3,1);
  \node [blue,above] at (3,1) {$\mathbf{e_{\alpha}}$};
  \node at (1,3.5) {fluid};
  \node at (4.5,1.5) {solid};
  %\draw [latex-latex] (2.95,1.05) --(2.45,1.55);
  %\node  at (2.5,1.1) {$q$};
\end{tikzpicture}
\caption{Curved interface on a square lattice : example of a fluid boundary node $\mathbf{x_f}$, its image in the solid domain $\mathbf{x_s}$, and the intersection point $ \mathbf{\displaystyle x_\Gamma}$ located on the interface}
  \label{fig:BBdraft}
\end{figure}

The fluid forces are computed using the momentum exchange method (MEM) proposed by Wen at al. \cite{WEN2014161}.
We note $\mathbf{x_f}$ a boundary node in the fluid domain and $\mathbf{x_s}$ the image of this boundary node through the solid interface by a lattice velocity $\mathbf{e_\alpha}$, also called incoming velocity( cf. figure \ref{fig:BBdraft}). The intersection point between the fluid-solid interface and the link $\mathbf{x_f} -\mathbf{x_s} $ is $\displaystyle \mathbf{\displaystyle x_\Gamma}$, and the outgoing lattice velocity is denoted $\displaystyle \mathbf{ \displaystyle e_{ \overline{\alpha}}} = - \mathbf{e_\alpha}$.
%and the first solid node $\mathbf{x_s}$,  $\mathbf{e_\alpha}$ denotes the incoming lattice velocity and $\mathbf{e_{\overline{\alpha}}}$ the opposite (figure \ref{fig:BBdraft}).
 
 The local force at $\mathbf{\displaystyle x_\Gamma}$ is computed using the following expression :
\begin{equation}\label{MMEL}
\mathbf{F} \left(\boldsymbol{\displaystyle x_\Gamma} \right) = \left(\mathbf{e_\alpha} - \mathbf{u_\Gamma} \right) \tilde{f}_\alpha \left( \boldsymbol{x_f}\right) -\left(\mathbf{e_{ \overline{\alpha}}} - \mathbf{u_\Gamma} \right) \tilde{f}_{ \overline{\alpha}} \left( \boldsymbol{x_s}\right)  
\end{equation}
and the total fluid force acting on the solid domain is :
\begin{equation}
\boldsymbol{\mathcal{F}_f}= \displaystyle \sum \mathbf{F} \left(\boldsymbol{\displaystyle x_\Gamma} \right)
\end{equation}

The torque is obtained with
\begin{equation}\label{fluidtorque}
\boldsymbol{\mathcal{T}_f}=\displaystyle \sum \left(\boldsymbol{x_\Gamma} -  \boldsymbol{x_G} \right) \times  \mathbf{F} \left(\boldsymbol{\displaystyle x_\Gamma} \right)
\end{equation}
Giovacchini and Ortiz \cite{PhysRevE.92.063302} showed that the MEM does not depend on the way the boundary conditions at the solid domain are implemented.
%The validity of the momentum exchange method for volume penalization approach can be shown using the approach of \cite{PhysRevE.92.063302}.

For each time step the fluid-structure problem is solved according to algorithm \ref{algo1}.

\begin{algorithm}
\caption{Calculate $\rho \left(t +\Delta t \right)$ and  $\mathbf{u} \left(t +\Delta t \right)$ \label{algo1}}
\begin{enumerate}
\item  Compute $\tilde{f}_\alpha \left(t\right)$, $f_\alpha \left(t +\Delta t \right)$
\item  Compute $ \mathcal{F}_f$ and $\mathcal{T}_f$
\item  Compute $\mathbf{x_G}\left(t +\Delta t \right)$ and $\boldsymbol{\theta}\left(t +\Delta t \right)$

A fourth-order Runge Kutta algorithm was used in this study

\item  Compute $\chi_{\Omega_s}\left(t +\Delta t \right) $ and $\mathbf{u_s}\left(t +\Delta t \right) $
\item Calculate  $\rho \left(t +\Delta t \right)$ and  $\mathbf{u} \left(t +\Delta t \right)$
\end{enumerate}
\end{algorithm}

In a previous work, Benamour et al. \cite{Benamour2015299} showed that the volume penalisation method combined with the LBM gives good results for computing flows around fixed bodies. In the following section, the VP-LBM is applied to different cases of fluid flows around moving bodies. 

\section{Numerical applications}
\label{sec:3}
The VP-LBM is applied to three cases. The first one considers the imposed transverse displacement of a cylinder in a fluid flow at a Reynolds number of $100$. The lift (and drag) coefficients computed with the VP-LBM are compared with those obtained using a classical CFD code (Code$\_$Saturne, \cite{codesaturne}). The second case is the study of particle sedimentation and the last one focuses on vortex-induced vibration (VIV) of a circular cylinder. 
All computations were run on a NVIDIA QUADRO P500 GPU card, using a CUDA implementation. For all computations, the following relaxation rates were chosen :
\begin{equation}
S = \displaystyle diag \left( \dfrac{1}{\tau}, 1.1,1.25,\dfrac{1}{\tau},1.8,\dfrac{1}{\tau},1.8,\dfrac{1}{\tau},\dfrac{1}{\tau}\right)
\end{equation}
where the relaxation time $\tau$ is related to the fluid viscosity thanks to the equation (\ref{tauvisco}), and a value of penalization factor $\eta = 10^{-6}$ was selected.

In the remaining of this paper $\textrm{l.u.}$ will refer to $\textrm{lattice length units}$ and $\textrm{t.s.}$ to $\textrm{lattice time units}$

%The refilling algorithm proposed by \cite{Tao20161} is used to recompute distribution functions missing on solid nodes which become fluid after the interface moves.
%This three methods are good candidates to evaluate the performances of the volume penalization methods.
% the Volume Penalisation consists only in adding a force term on solid nodes, and in the implemented version of this study does not need the computation of the distance between nodes and interface.

%The Bounce-Back method implemented in this study is these proposed by \cite{Bouzidi20013452}, and the refilling algorithm to extrapolate values where a node change of domain comes from \cite{lallemand_lattice_2003}. The Lift and Drag coefficients are computed using the momentum exchange method  \cite{Bouzidi20013452}.

\subsection{Imposed displacement of a cylinder in a transverse fluid flow \label{sec:app1}}

\begin{figure}[!htpb]
  \centering
  \begin{tikzpicture}
%    \draw [help lines] (0,0) grid (12,5); 
    \draw [thick] (0,0) -- (12,0);
 \draw [thick] (0,5) -- (12,5);
 \draw [thick] (3,2.5) circle [radius=0.5];
% \draw (0,0) .. controls (1,2) and (1,3) .. (0,5);
%spring
 \draw (3,2) --(3,1.87)-- (3.25,1.75) -- (2.75,1.50) --(3.25,1.25)--(2.75,1) -- (3.25,0.75) -- (3,0.63)--(3,0.5) ;
\draw (2.75,0.5)-- (3.25,0.5);
\draw (2.75,0.5) -- (2.65,0.4);
\draw (2.85,0.5) -- (2.75,0.4);
\draw (2.95,0.5) -- (2.85,0.4);
\draw (3.05,0.5) -- (2.95,0.4);
\draw (3.15,0.5) -- (3.05,0.4);
\draw (3.25,0.5) -- (3.15,0.4);
%inlet
\draw [cyan,thick,-latex] (0,0) -- (0.5,0);
\draw [cyan,thick,-latex] (0,0.5) -- (0.5,0.5);
\draw [cyan,thick,-latex] (0,1) -- (0.5,1);
\draw [cyan, thick,-latex] (0,1.5) -- (0.5,1.5);
\draw [cyan, thick,-latex] (0,2) -- (0.5,2);
\draw [cyan, thick,-latex] (0,2.5) -- (0.5,2.5);
\draw [cyan,thick,-latex] (0,3) -- (0.5,3);
\draw [cyan,thick,-latex] (0,3.5) -- (0.5,3.5);
\draw [cyan,thick,-latex] (0,4) -- (0.5,4);
\draw [cyan,thick,-latex] (0,4.5) -- (0.5,4.5);
\draw [cyan,thick,-latex] (0,5) -- (0.5,5);
% Dimension
\draw [latex-latex] (2.65,2.15) --(3.35,2.85); % -- (4,2.85);
\draw (3.35,2.85) -- (4,2.85);
\node [right] at (4,2.85) {D};
\draw [latex-latex] (0,1.75) -- (3,1.75);
\node [below] at (1.5,1.75) {$L_1$};
\draw [latex-latex] (3,1.75) -- (12,1.75);
\node [below] at (7.5,1.75) {$L_2$};
\draw [latex-latex] (12.5,0) -- (12.5,5);
\node [right] at (12.5,2.5) {$H$};
% BC
\node [left] at (0,2.5) {$U_0$};
\node [left] at (0,3) {inlet};
\node [left] at (12,3) {outlet};
\node [left] at (8,3) {symmetry};
\draw [-latex] (7,3.5) .. controls (7.1,4) and (6.75,4.45) .. (6,5);
\draw [-latex] (7,2.5) .. controls (7.1,2) and (7,1) .. (6,0);

%repere
\draw [latex-latex] (1.5,4.5)--(1.5,3.5) -- (2.5,3.5);
\node [above] at (2.5,3.5) {$\mathbf{x}$};
\node [right] at (1.5,4.5) {$\mathbf{y}$};
\end{tikzpicture}
  \caption{Imposed displacement of a cylinder in a transverse flow}
  \label{fig:cylindre}
\end{figure}
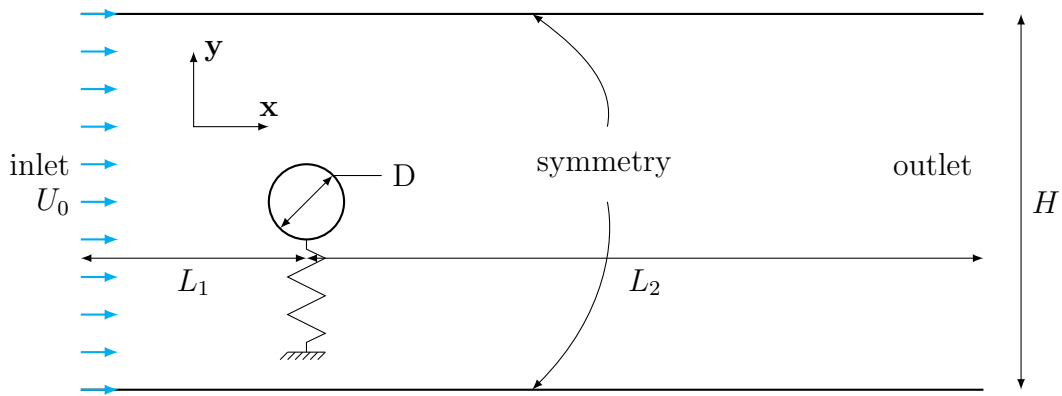
To validate the VP-LBM, the imposed displacement of a cylinder in a transverse flow at a Reynolds number of $100$ was first simulated (figure (\ref{fig:cylindre})). 
For that case, a constant velocity profile was imposed at the inlet using the classical half-way Bounce-Back method, and the outflow boundary condition at outlet was modelled using the convective condition \cite{Yang2013160}. Symmetry boundary conditions ( $\mathbf{u}\cdot \mathbf{n}$) were imposed at the other boundaries, in order to apply the same boundary conditions as those employed with the finite volume code (Code$\_$Saturne).

 The computational procedure is the following : first, a computation was run without solid displacement to obtain a well-established fluid flow. This state is considered as the initial time $\left(t=0 \right)$. Next, the motion of the cylinder was imposed according to the following expression :
\begin{equation}
  y_G\left( t \right)=A +B\textrm{cos}\left( \omega t\right)=\left( y_G\left( 0 \right) -\dfrac{D}{4} \right) + \dfrac{D}{4}  \textrm{cos}\left( \omega t\right)
 \label{eq:displacementimposed}
\end{equation}
where $y_G$ is the $\mathbf{y}$ coordinate of the center of gravity, and $D$ the cylinder diameter, and the fluid flow was calculated.

To test the ability of the VP-LBM to calculate flows around moving obstacles, and to validate the computation of the hydrodynamic force  (equation (\ref{MMEL})) exerted on the moving body, the results are compared with those obtained by a conventional computational code, the finite volume software Code$\_$Saturne \cite{codesaturne}. To compare the results obtained with the VP-LBM, and with the finite volume method, the same non dimensional numbers were used :
\begin{equation}
  \omega^\star=\omega\dfrac{D}{U_0} \hspace{1cm}  \textrm{B}^\star=\dfrac{\textrm{B}}{D} \hspace{1cm} \mathbf{u}^\star= \dfrac{\mathbf{u}}{U_0}
  \label{eq:imposedad}
\end{equation}
In this study: $ \omega^\star=1.55 $, and the LBM parameters (in lattice units) are 
$$\tau = 0.56 \hspace{0.5cm} U_0=0.048780  \;\textrm{l.u.}/\textrm{t.s.}  \hspace{0.5cm}  L_1 =1230\;\textrm{l.u.}  \hspace{0.5cm} L_2= 410 \;\textrm{l.u.}    \hspace{0.5cm} D=41 \;\textrm{l.u.} $$ 
\begin{figure}[!htbp]
\center
\subfigure[  Grid used for the finite volume computation (Code$\_$Saturne) \label{meshsat}]{\includegraphics[width=5cm]{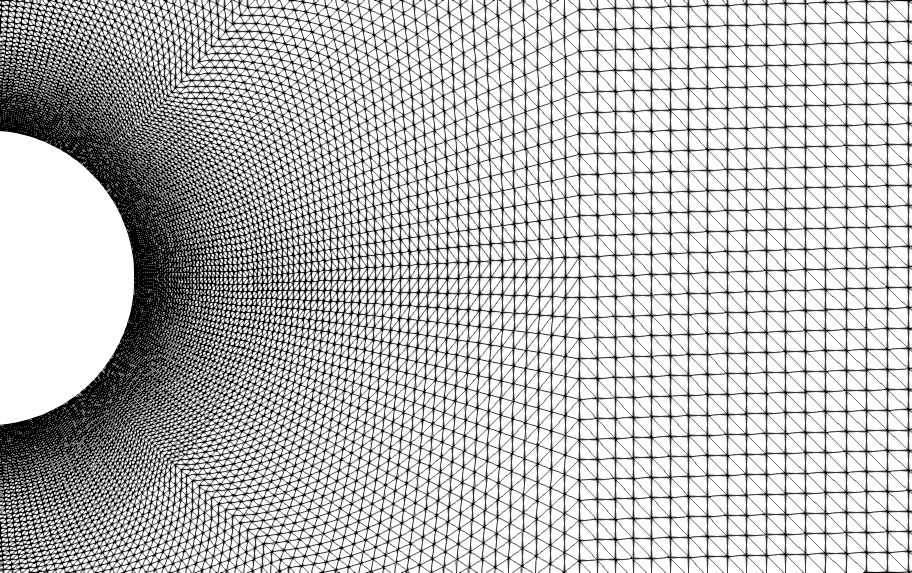}}
\subfigure[Characteristic function used for the VP-LBM \label{fcar}]{\includegraphics[width=5cm]{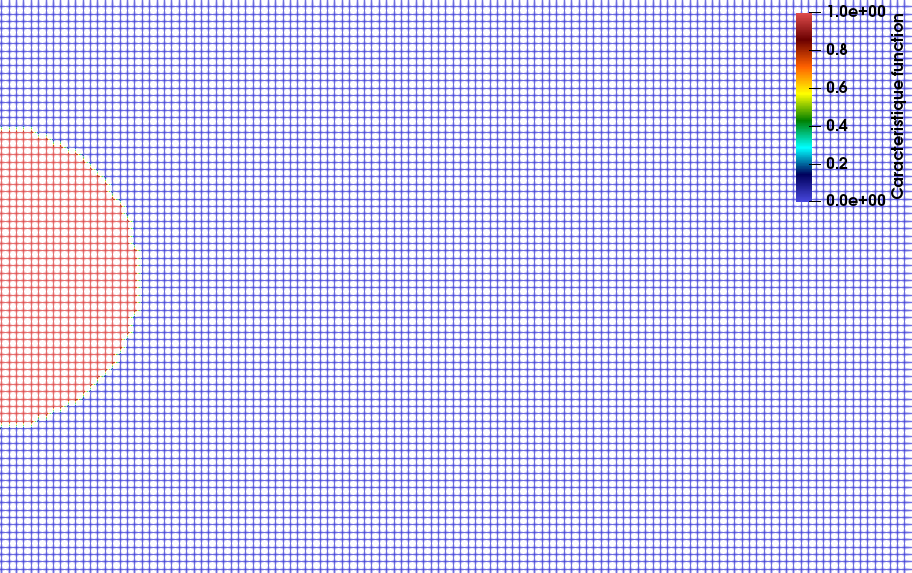}}
\caption{ Grids used for the finite volume computations and for the VP-LBM computations \label{meshfcar}}
\end{figure}

For the computation performed with Code$\_$Saturne, a non uniform grid that was very fine in the vicinity of the cylinder, and coarser in the remaining of the fluid domain, was built (see figure \ref{meshsat}). Such a mesh enabled a decrease in computational time. For the LBM computations, a regular grid was used.
 
\begin{figure}[!htbp]
\center
\subfigure[ $\Vert \mathbf{u}^\star \Vert $ Code$\_$Saturne  \label{compcfd}]{\includegraphics[width=5cm]{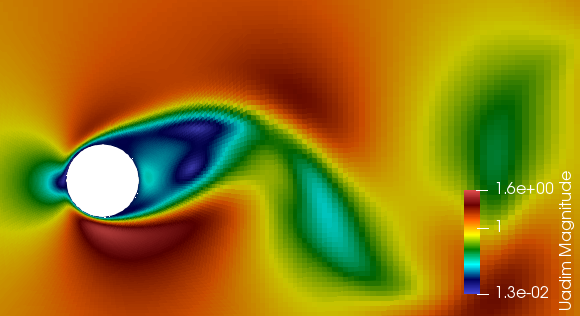}}
\subfigure[$\Vert \mathbf{u}^\star \Vert $  VP-LBM \label{complbm}]{\includegraphics[width=5cm]{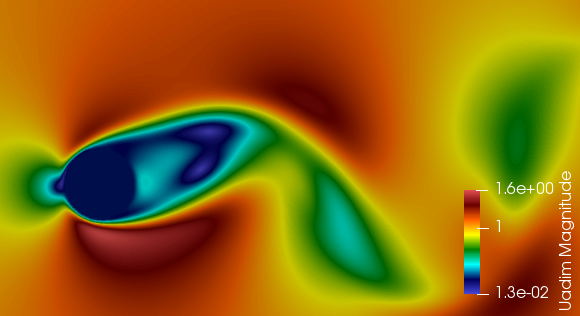}}
\caption{Non-dimensional velocity fields obtained with Code$\_Saturne$, and with the VP-LBM computations, for the imposed motion of a cylinder \label{comp1}}
\end{figure}

Figure \ref{comp1} shows the norms of the non-dimensional velocity field computed with Code$\_$Saturne software and with the VP-LBM at the same non-dimensional time $t^\star = 64$, where $t^\star =t \dfrac{U_0}{D}$. It can bee seen that both computations yield the same velocity field. More particularly, the vortex which appears behind the moving cylinder is located at the same place.
%Far from the cylinder the LBM grid is finer than those used with Code$\_$Saturne which explain th

\begin{figure}[!htbp]
	\centering
	\subfigure[$C_L$]{\includegraphics[width=6cm]{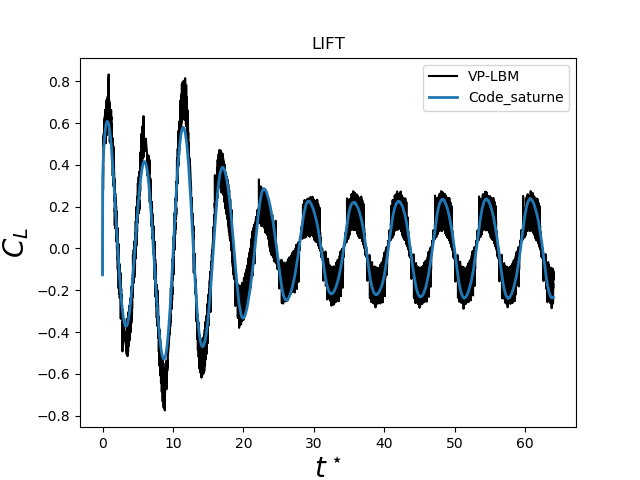}}
	\subfigure[$C_D$]{\includegraphics[width=6cm]{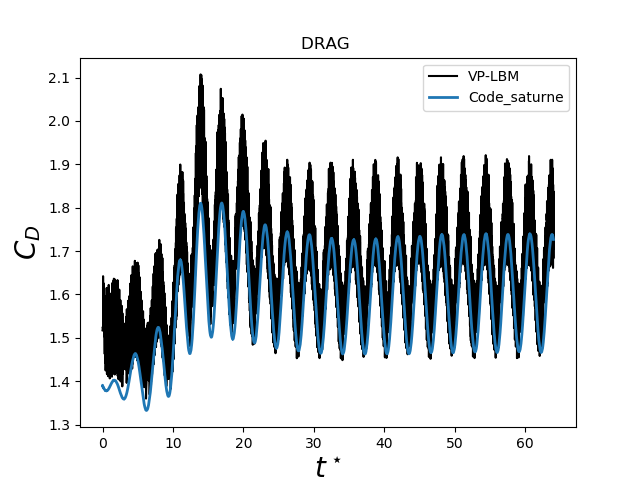}}
\caption{Comparison of $C_L$ and $C_D$ obtained with the VP-LBM and Code$\_$Saturne  }
	\label{fig:cas1compR0}
      \end{figure}

In figures \ref{fig:cas1compR0}, the lift $\left(C_L  \right)$ and drag $\left(C_D \right)$ coefficients are compared :
$$
C_L= \dfrac{\boldsymbol{\mathcal{F}_f}\cdot \mathbf{y}}{\dfrac{1}{2}\rho D \left(U_0\right)^2 } \hspace{1cm} C_D= \dfrac{\boldsymbol{\mathcal{F}_f}\cdot \mathbf{x}}{ \dfrac{1}{2}\rho D \left(U_0\right)^2}
$$
In this figure, it can be noticed that spurious oscillations occur when LBM is used. However, when using LBM, a cartesian grid is commonly used, and this behavior has been highlighted for computations of FSI problems carried out with LBM, combined with any method chosen for simulating flows around curved boundaries  (Bounce-back or immersed boundary methods) (\cite{WEN2014161}). A good agreement between both methods was obtained for the lift coefficient, whereas small differences are observed for the drag coefficient. These differences are not discriminating for the VP-LBM approach, because as shown in table \ref{tabcd}, the difference based on the average drag is less than $5 \%$, and this gap is in the range of what can be expected when two different numerical models are compared.

\begin{table}[!htbp]
\center
\begin{tabular}{|c|c|c|}
\hline 
• & VP-LBM & Code\_Saturne \\ 
\hline 
$\overline{C_D}$ & 1.653 & 1.577 \\ 
\hline 
\end{tabular} 
\caption{\label{tabcd} Average drag coefficient obtained with the VP-LBM and Code$\_$Saturne}
\end{table}

To conclude with this first example, the VP-LBM predicts accurately the fluid forces exerted on a solid whose motion is imposed. In order to test the validity of the VP-LB, the following examples deal with real cases of fluid structure interaction.

\paragraph{Remark } In order to reduce the size of the LBM problem ($L_1 \times L_2$) and thus the computational time, the value of the parameter $\tau=0.56$ was chosen very close to the stability limit $0.5$. An inlet velocity of $0.048780$ ensured a small Mach number suitable for the LBM approach. For that simulation, $41$ lattice units were used for the cylinder diameter, which is a small value for computing a flow of Reynolds number $100$ with LBM. This can explain the spurious oscillations. 

\subsection{Sedimentation of a particle under gravity}

\begin{figure}[!htpb]
%\begin{minipage}{0.48\textwidth}
  \centering
  \begin{tikzpicture}
%    \draw [help lines] (0,0) grid (12,5); 
    \draw [line width=5 pt] (0,0) -- (0,4);
 \draw [line width=5 pt] (3.8,0) -- (3.8,4);
 \draw [thick] (1.5,3) circle [radius=0.5];
% \draw (0,0) .. controls (1,2) and (1,3) .. (0,5);
% Dimension
\draw [right,latex-latex] (4,0) --(4,4);
\node [right] at (4,2) {L};
\draw [below,latex-latex] (0,-0.1) -- (3.8,-0.1);
\node [below] at (2,0) {$H$};
\draw [latex-latex] (1.1,2.7) -- (1.9,3.27);% -- (2.5,3.28);
\draw (1.9,3.27) -- (2.5,3.28);
\node [above] at (2.5,3.28) {D};
%repere
\draw [latex-latex] (-1.5,1.5)--(-1.5,0.5) -- (-0.5,0.5);
\node [above] at (-0.5,0.5) {$\mathbf{x}$};
\node [right] at (-1.5,1.5) {$\mathbf{y}$};
%_
\draw [latex-latex] (0,2.4) -- (1.5,2.4);
\node [below] at (0.75,2.5) {$x_0$};
\draw [latex-latex] (0.9,3) -- (0.9,4);
\node [left] at (0.9,3.5) {$y_0$};

\draw [red,line width=2 pt,-latex] (3,3) -- (3,2);
\node [right] at (3,2.5) {$\boldsymbol{g}$};
\end{tikzpicture}
  %\caption{sedimentation configuration}
%\end{minipage}
%\begin{minipage}{0.48 \textwidth}
%  \begin{itemize}
%    \item $\rho_f=1 \; \textrm{g}\cdot \textrm{cm}^{-3}$
%    \item $D = 0.24 \; \textrm{cm}$
%    \item $\nu= 0.1 \; \textrm{g}\cdot\textrm{cm}^{-1}\cdot \textrm{s}^{-1}$ 
%    \item $g=980 \; \textrm{cm}\cdot\textrm{s}^{-2}$
%  \end{itemize}
%\end{minipage}
\caption{Schematic description of particle sedimentation}
  \label{schemsuspension}
\end{figure}
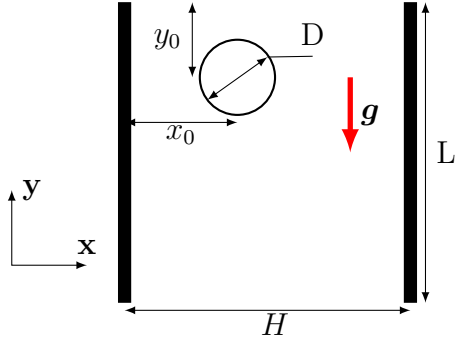

The next case focuses on the sedimentation of a particle under gravity in an infinite channel (figure \ref{schemsuspension}) for centered and non-centered configurations. These problems have been extensively used for model validations and are very useful to test the capacity of a method to capture complex trajectories \cite{Wang20131151,Tao20161,WEN2014161}.

A circular particle of diameter $D$ falls under gravity $\boldsymbol{g}$ in a fluid of density $\rho$ in a vertical channel of width $H$. At the initial state, the particle is at a distance $x_0$ from the left wall, a distance  $y_0$ from the top of the channel and the velocity of the particle is equal to zero. For that case, the particle displacement can be described using equations (\ref{xstru} ) and (\ref{thetastru}) modified as follows :
\begin{eqnarray}
%\begin{array}{ll}
m \dfrac{d^2 \boldsymbol{x_G}}{d t^2} &=&\boldsymbol{\mathcal{F}_f} + m\left( 1 - \dfrac{\rho}{\rho_s} \right) \boldsymbol{g} \label{dep1}\\
I \dfrac{d^2 \boldsymbol{\theta}}{d t^2}  &=& \boldsymbol{\mathcal{T}_f} \label{theta2}
%\end{array}
\end{eqnarray}
where $\rho$ denotes the fluid density and $\rho_s$ the solid density. The last term in equation (\ref{dep1}) represents the weight and the buoyancy (Archimedes' principle) acting on the particle.
%\begin{equation}
%\left\{
%  \begin{array}{ll}
%    \dfrac{d \mathbf{x_G}}{d t} = \mathbf{v_p} \\
%    m_p\dfrac{ d \mathbf{v_p}}{d t} = \mathcal{F}_f + m_p\left( 1 - \dfrac{\rho_f}{\rho_p} \right) \boldsymbol{g}
%  \end{array}\right.
%  \label{eq:motionparticule}
%\end{equation}
%where $\mathbf{x_G}$ denotes the vector position of the gravity center of the particle, $\mathbf{v_p}$ is the particle velocity, $m_p$ is its mass, $\mathcal{F}_f$ are the fluid forces, $\rho_p$ is the particle density and $\boldsymbol{g}$ is the gravity vector.

For small Reynolds numbers and a large non dimensional width $\tilde{H}=\dfrac{H}{D}$, the particle reaches a steady state, where the drag force can be approximated according to the following expression \cite{happel}:
\begin{equation}
  \mathcal{F}_D=\dfrac{1}{4} \pi D^2 \left( \rho_p-\rho_f \right)=4 \pi \kappa \mu \mathbf{v_s}\cdot\mathbf{y}
  \label{exp:dragforces}
\end{equation}
where $\kappa$ denotes the correction factor which represents the channel confinement effect,
\begin{equation}
\kappa = \displaystyle \left( \textrm{ln}\tilde{H}-0.9157 + 1.7244 \tilde{H}^{-2} - 1.7302 \tilde{H}^{-4} + 2.4056 \tilde{H}^{-6} - 4.5913 \tilde{H}^{-8} \right)^{-1}
\end{equation}
 $\mu$ is the fluid viscosity and $\mathbf{v_s}$ is the gravitational settling of the particle :
\begin{equation}
  \mathbf{v_s}=\dfrac{D^2}{16\kappa \mu} \left( \rho - \rho_s\right) \boldsymbol{g}
  \label{exp:velocitysteady}
\end{equation}
\subsubsection{Centered particle : $x_0= \dfrac{H}{2}$}
%First, the case where $x_0= \dfrac{H}{2}$ is considered. 
%  \begin{column}{0.48\textwidth}    

We consider the same physical properties as those  chosen in previous works \cite{Wang20131151, Dorschner2015340}. The diameter of the particle is $D=0.24\; \textrm{cm}$, the fluid density is $\rho=1 \;\textrm{g}\cdot\textrm{cm}^{-3}$, the fluid viscosity is $\mu=0.1 \; \textrm{g}\cdot \textrm{cm}^{-1} \cdot\textrm{s}^{-1}$, the gravitational acceleration is $\Vert  \boldsymbol{g} \Vert = 980 \; \textrm{cm} \cdot \textrm{s}^{-2}$ and the non dimensional width of the channel is $\tilde{H}=5$.

The problem was modelled with a regular $120\times 1200$ lattice grid, $24$ nodes across the diameter of the cylinder ($D_{LBM}=24 \; \textrm{l.u.}$), and a relaxation time $\tau = 0.8$. Initially, the particle is located at the center of the channel : $x_0=0.5 H$ and $y_0=0.5L$. No-slip boundary conditions were imposed on the left and right walls. A zero velocity boundary condition was applied at the inlet (top of the channel) and free stream conditions were applied at the outlet (bottom). A large value of $L$ was chosen, so that the inlet and outlet did not influence the behavior of the particle.

Computations were performed for various mass ratios $\rho_r=\dfrac{\rho_s}{\rho}$ taken from the following list :
\begin{equation}
  0.95; 0.98; 0.99; 1.01; 1.02; 1.05 
  \label{listmassnumber}
\end{equation}

%All LBM simulation are carried out with the same relaxation time ($\tau=0.8$).
%the other LBM parameters ($\boldsymbol{V}_{LBM}$) is computed using the estimated Reynolds number using the velocity expressed by expression \ref{exp:velocitysteady}. At least, the gravity value introduced in the Boltzmann model is computed using the following equation :
%\begin{equation}
%  g_{LBM}=\dfrac{\boldsymbol{V}_{LBM}}{\dfrac{D^2_{LBM}}{16 \kappa \nu_{LBM}}\left( 1-\rho_r \right)}
%  \label{glbm}
%\end{equation}

%An explicit coupling algorithm is used to solve the FSI problem. The fluid forces are solved using the same method as in the previous example. 

\begin{figure}[!htbp]
  \centering
  \subfigure[Particle velocities]{\includegraphics[width=6cm]{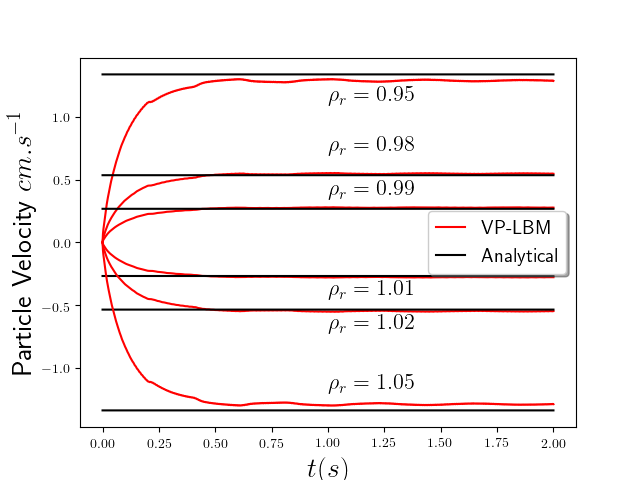}}
  \subfigure[Drag coefficients]{\includegraphics[width=6cm]{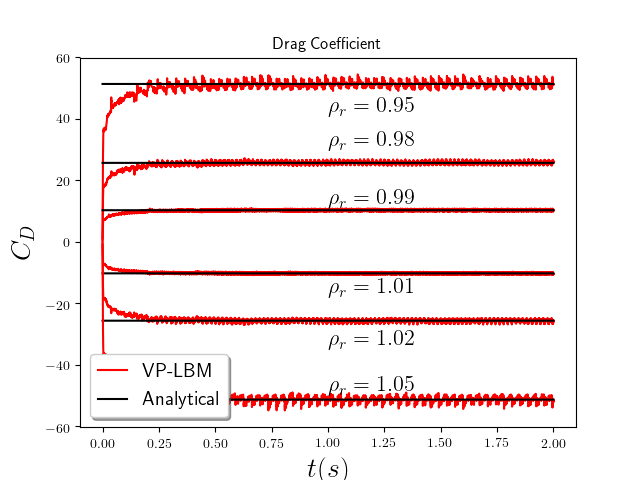}}
  \caption{Comparison of particle velocities and drag coefficients for different mass ratios (Centered particle and $\tilde{H}=5$)}
  \label{fig:resususpen1}
\end{figure}

For all cases  the particle velocities and the drag coefficients are plotted in figure \ref{fig:resususpen1} and compared with the analytical solution (equations (\ref{exp:dragforces}) and (\ref{exp:velocitysteady})).

For $\rho_r=\left\{0.98; 0.99; 1.01; 1.02\right\}$, results match with the analytical solution. For larger ratios ($\rho_r=\left\{0.95; 1.05\right\}$) differences are observed. As mentioned in previous studies \cite{Tao20161,Dorschner2015340}, the reason could be that the analytical solutions are available only for small mass numbers $\rho_r$ and not for larger ratios such as $1.05$ or $0.95$, which explains the gap between the analytical solution and the results obtained with the VP-LBM approach. As noticed in the previous example, small fluctuations on the drag coefficient are observed when the velocity increases $\left(\rho_r=0.95 \; \textrm{or} \; 1.05 \right)$, but the averages fit with the analytical solutions. 

%
%\begin{figure}[!htbp]
%  \centering
%  \subfigure[$\rho_r=0.95$]{\includegraphics[width=6cm]{figure/v95.png}}
%  \subfigure[$\rho_r=102$]{\includegraphics[width=6cm]{figure/v102.png}}
%  \caption{Particle velocity magnitude at time $t=2\; \textrm{s}$ for $\rho_r=0.95$ and $1.02$}
%  \label{fig:vmagncp}
%\end{figure}

%This example at small Reynolds numbers validates the idea of combining the Volume Penalization method with LBM for computing fluid structure interaction %problems. The results that we obtained are in the same order a those of existing methods. The following example focuses on a classical Vortex Induced Vibration %(VIV) case. The capacity of the VP-LBM method will be tested for complex flow and coupling.  
\subsubsection{Non centered particle : $x_0=0.75 D$}
The following case deals with a particle whose initial position is not at the center of the channel. The fluid properties are  $\rho=1 \;\textrm{g}\cdot\textrm{cm}^{-3}$, and $\mu=0.1 \; \textrm{g}\cdot \textrm{cm}^{-1} \cdot\textrm{s}^{-1}$ and the physical problem concerns a particle of diameter $D=0.1 \; \textrm{cm}$, a mass ratio $\rho_r = 1.03$ and $\Vert  \boldsymbol{g} \Vert = 980 \; \textrm{cm} \cdot \textrm{s}^{-2}$. The Reynolds number based on the final velocity of the particle is $Re= 8.33$.

For the LBM computations a $125 \times 1550$ lattice grid was used, the cylinder diameter was $31 \; \textrm{l.u.}$, the non dimensional relaxation time was $\tau=0.6$ and the particle was released at $y_0=12.5D$. The boundary conditions were the same as in the previous case.

\begin{figure}[!htbp]
\center
\subfigure[position of the particle]{\includegraphics[width=6cm]{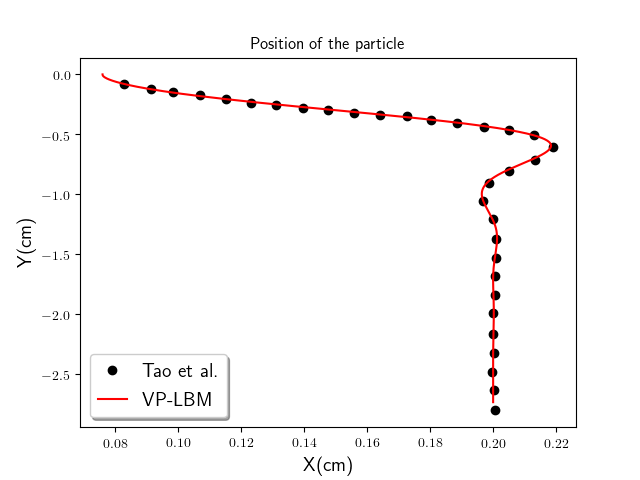}}
\subfigure[horizontal velocity]{\includegraphics[width=6cm]{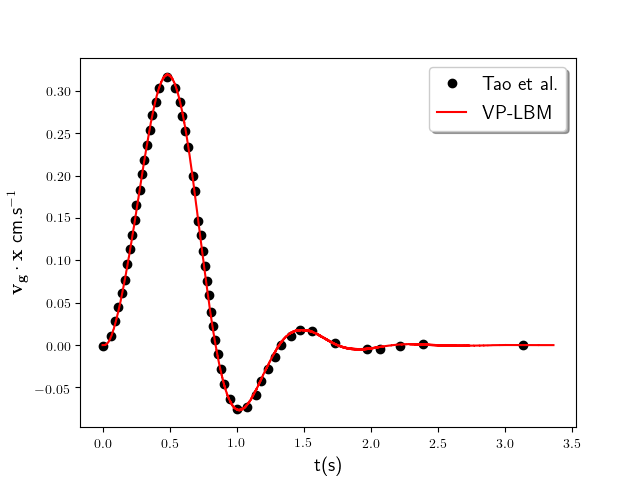}}
\subfigure[vertical velocity]{\includegraphics[width=6cm]{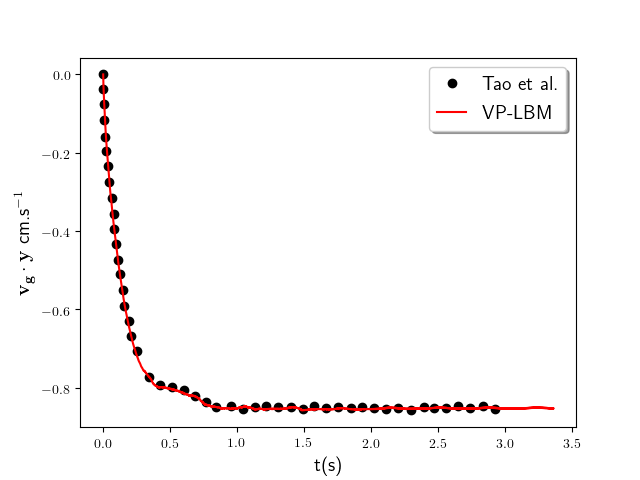}}
\subfigure[\label{fig:rv} rotational velocity]{\includegraphics[width=6cm]{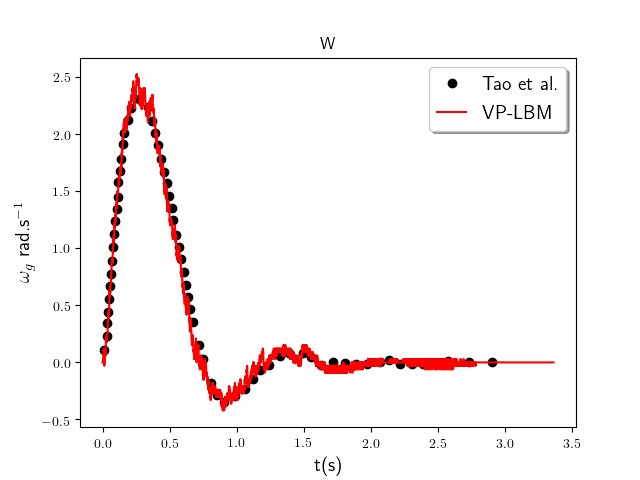}}
\caption{\label{ncparticle1} Results obtained using the VP-LBM apporach and compared with Tao et al's results \cite{Tao20161}}
\end{figure}

Results are plotted in figures \ref{ncparticle1}. $\mathbf{v_g}$ denotes the vector velocity of the gravity center of the particle, and $\omega_g$ its rotational velocity. The time-dependent position, horizontal, vertical and rotational velocities are compared with results from Tao et al. \cite{Tao20161} and good agreements are found between them. Small spurious oscillations are observed for the rotational velocity (figure \ref{fig:rv}), but these oscillations are also observed by Tao et al. \cite{Tao20161}. 

\begin{figure}[!htbp]
\center
\subfigure[\label{u1} t = 0.4 s]{\includegraphics[width=3cm]{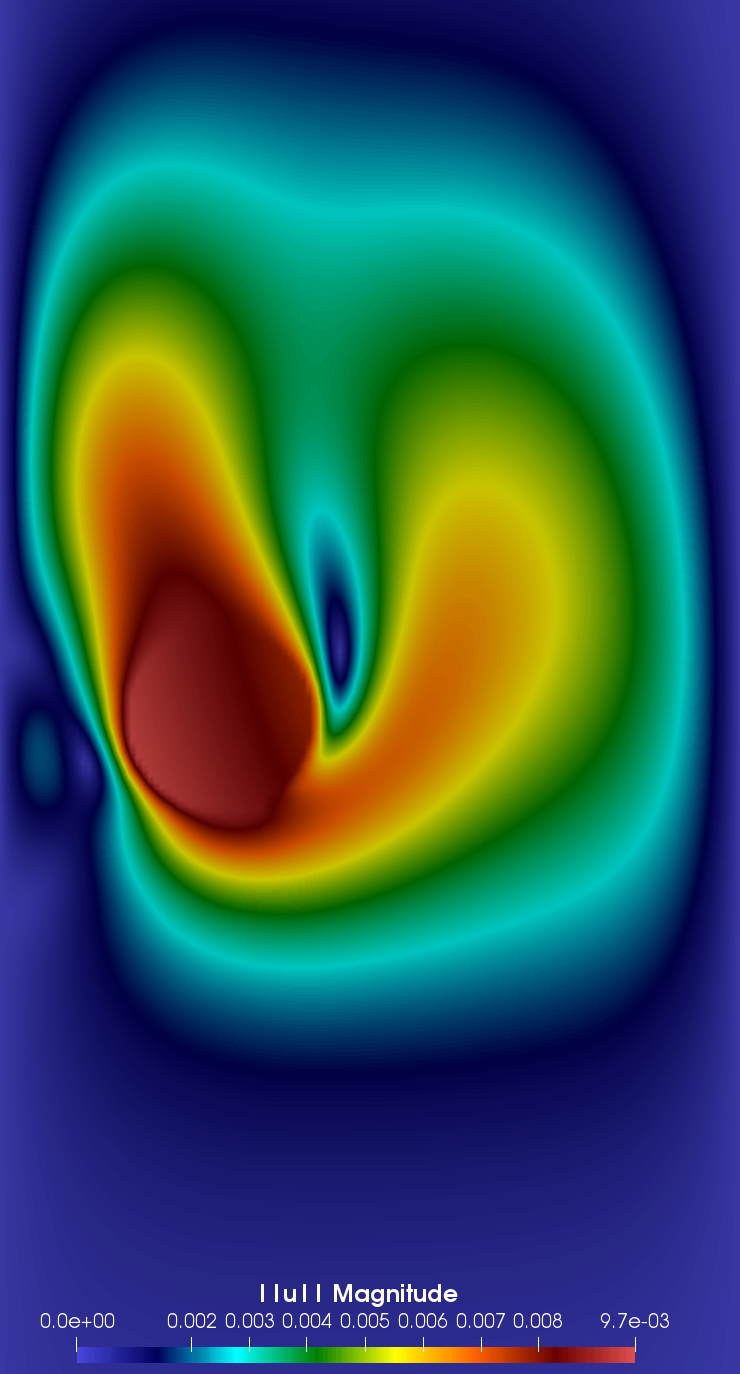}}
\subfigure[\label{u2} t = 0.6 s]{\includegraphics[width=3cm]{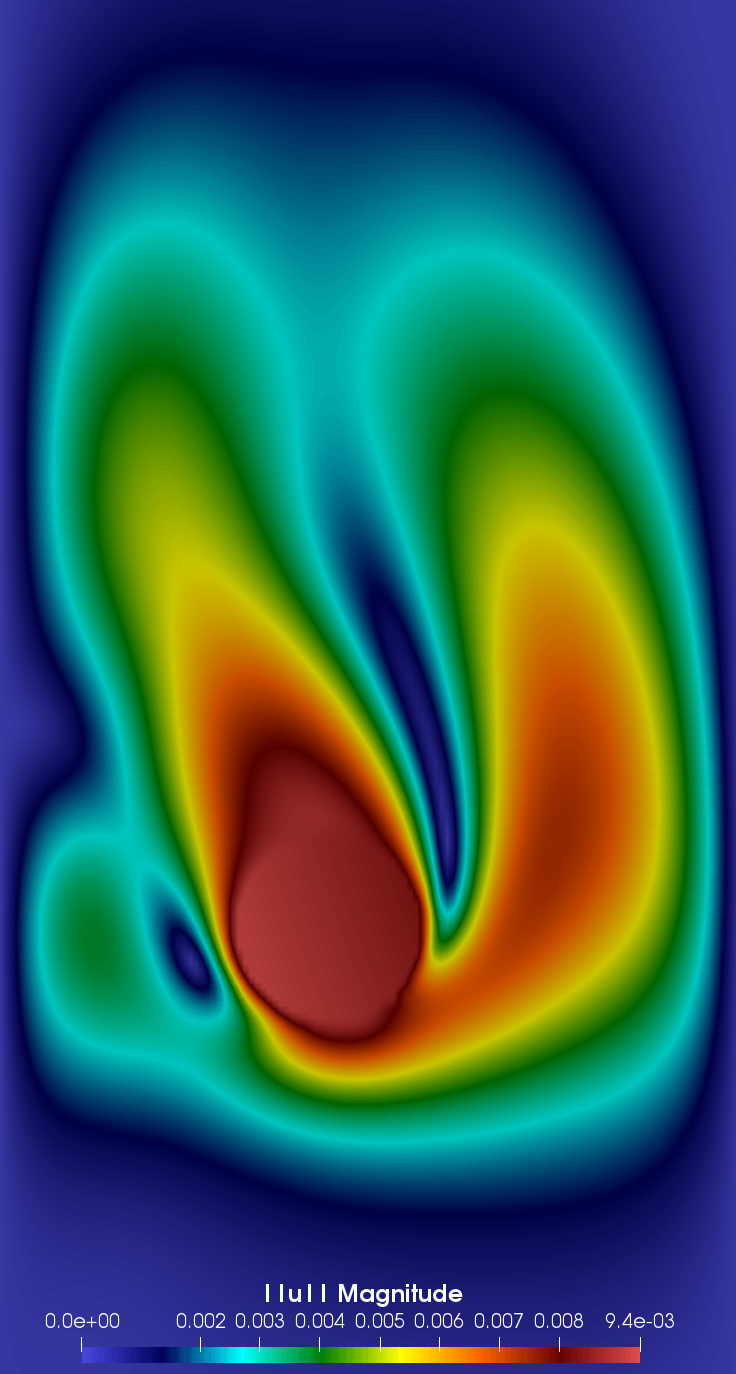}}
\subfigure[\label{u3} t = 1.0 s]{\includegraphics[width=3cm]{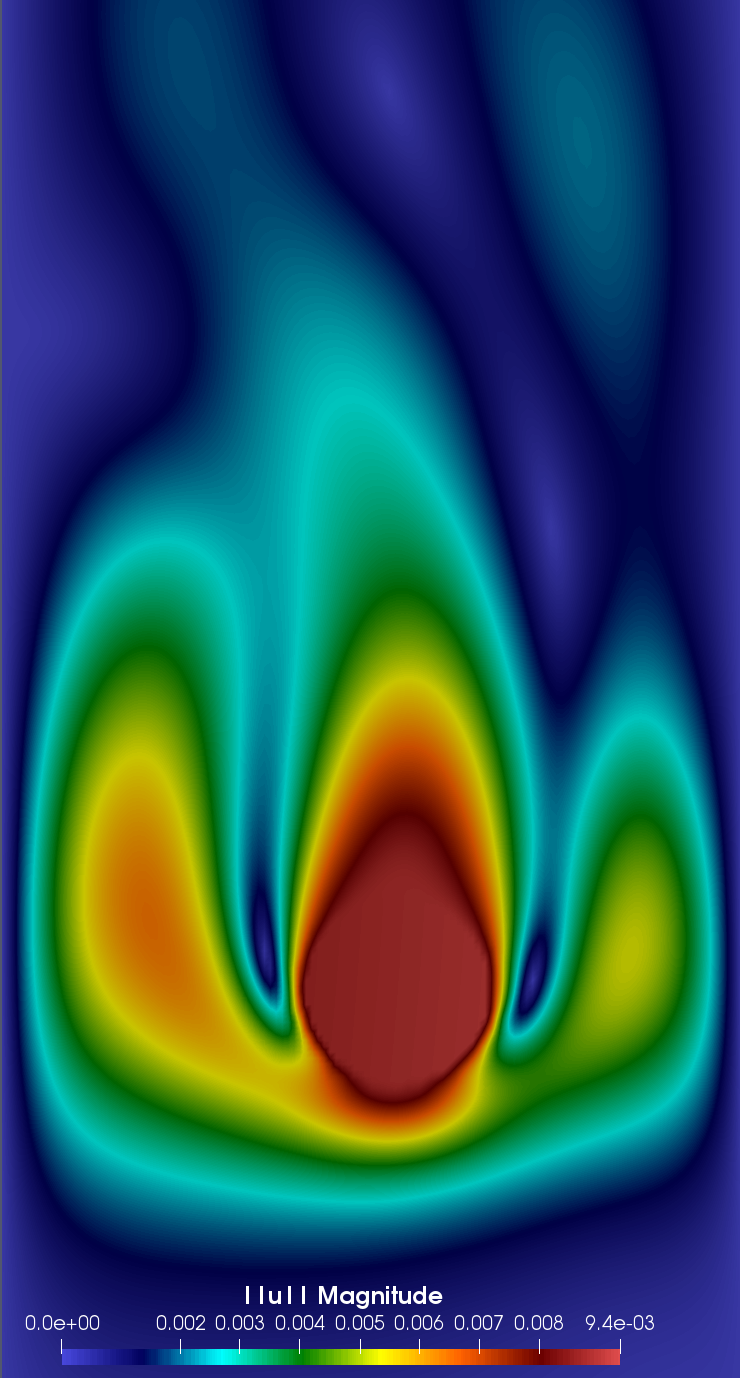}}
\subfigure[\label{u4} t = 3.0 s]{\includegraphics[width=3cm]{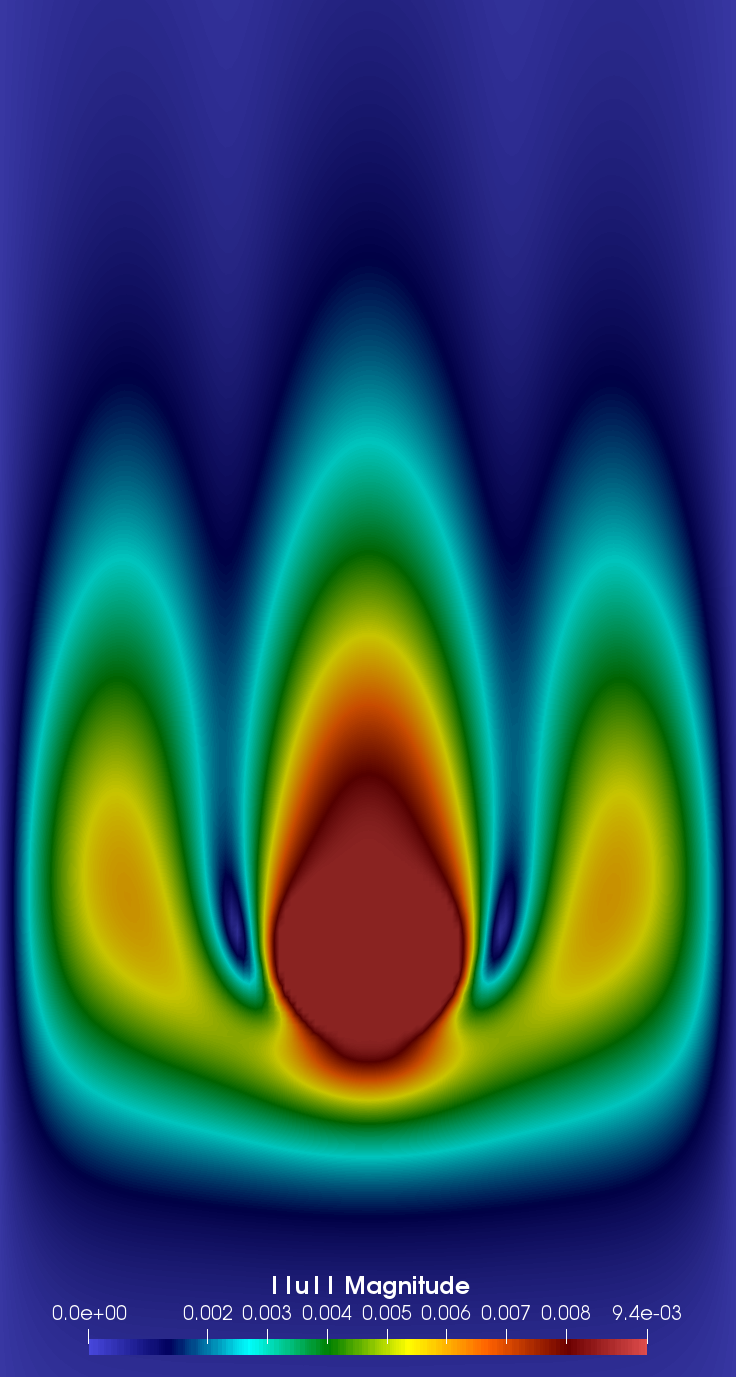}}
\caption{\label{vnc} Fluid velocity magnitude at times t=0.4, 06, 1.0 and 3.0 seconds in lattice units}
\end{figure}

\begin{figure}[!htbp]
\center
\subfigure[\label{v1} t = 0.4 s]{\includegraphics[width=3cm]{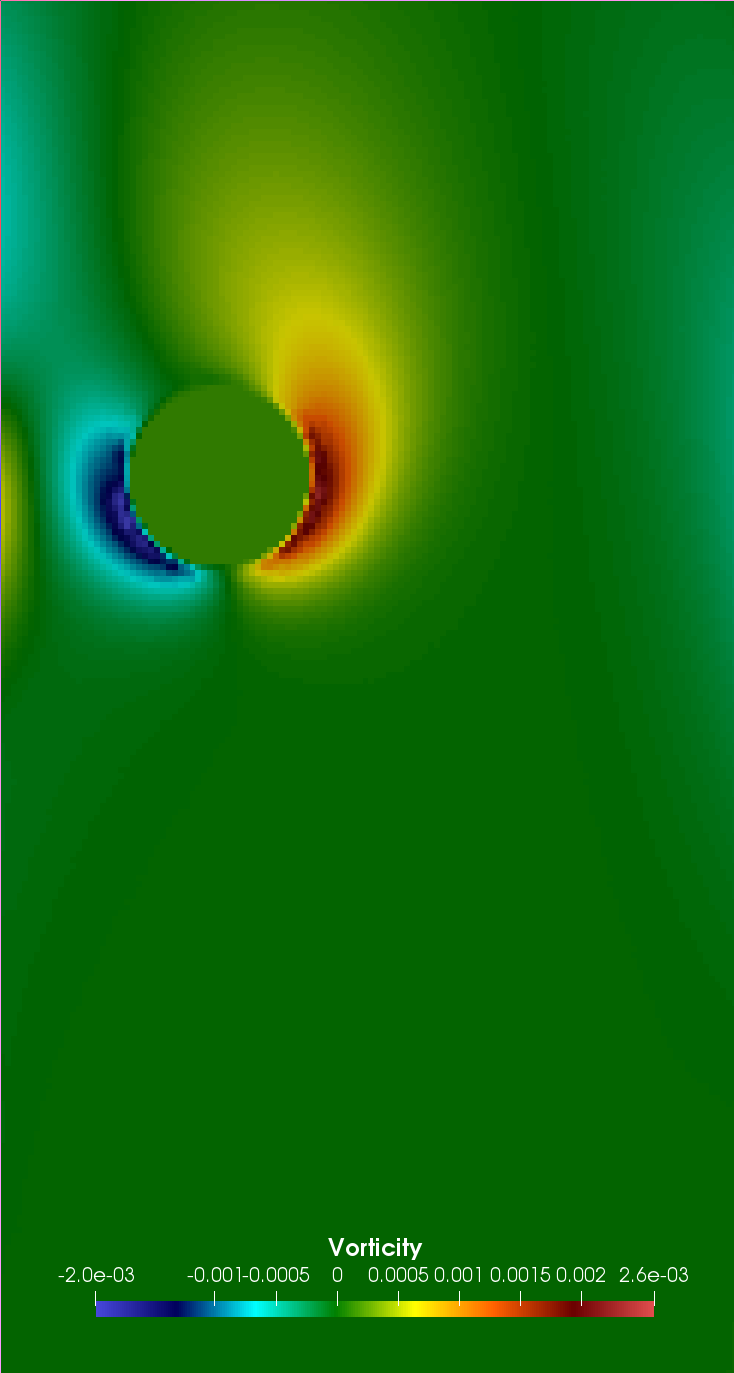}}
\subfigure[\label{v2} t = 0.6 s]{\includegraphics[width=3cm]{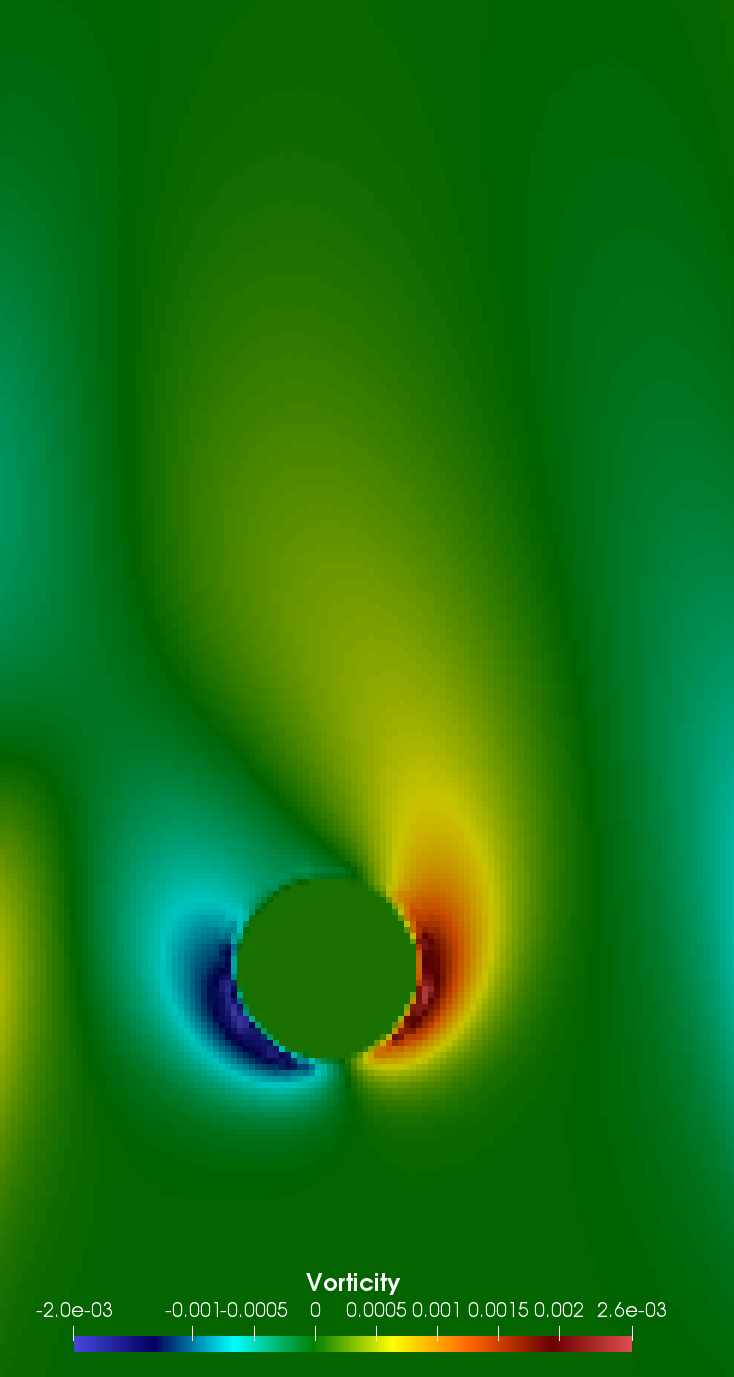}}
\subfigure[\label{v3} t = 1.0 s]{\includegraphics[width=3cm]{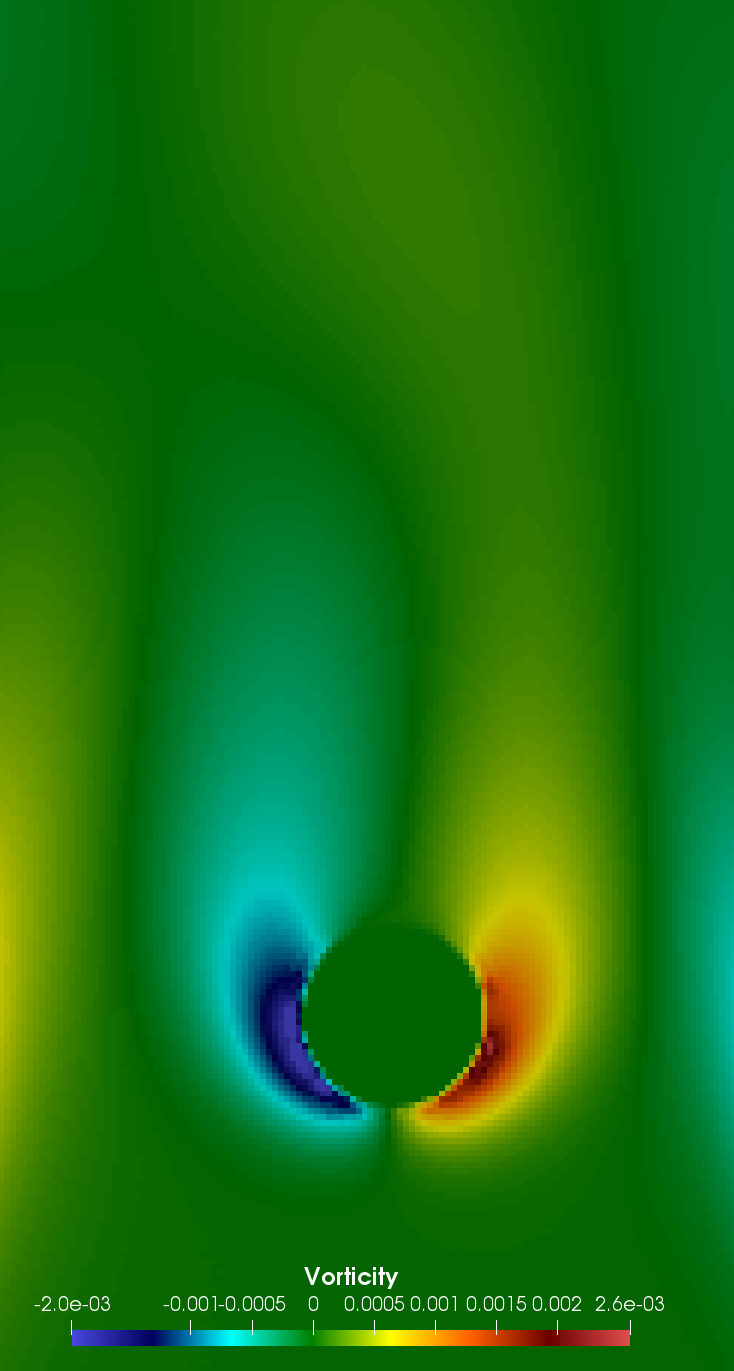}}
\subfigure[\label{v4} t = 3.0 s]{\includegraphics[width=3cm]{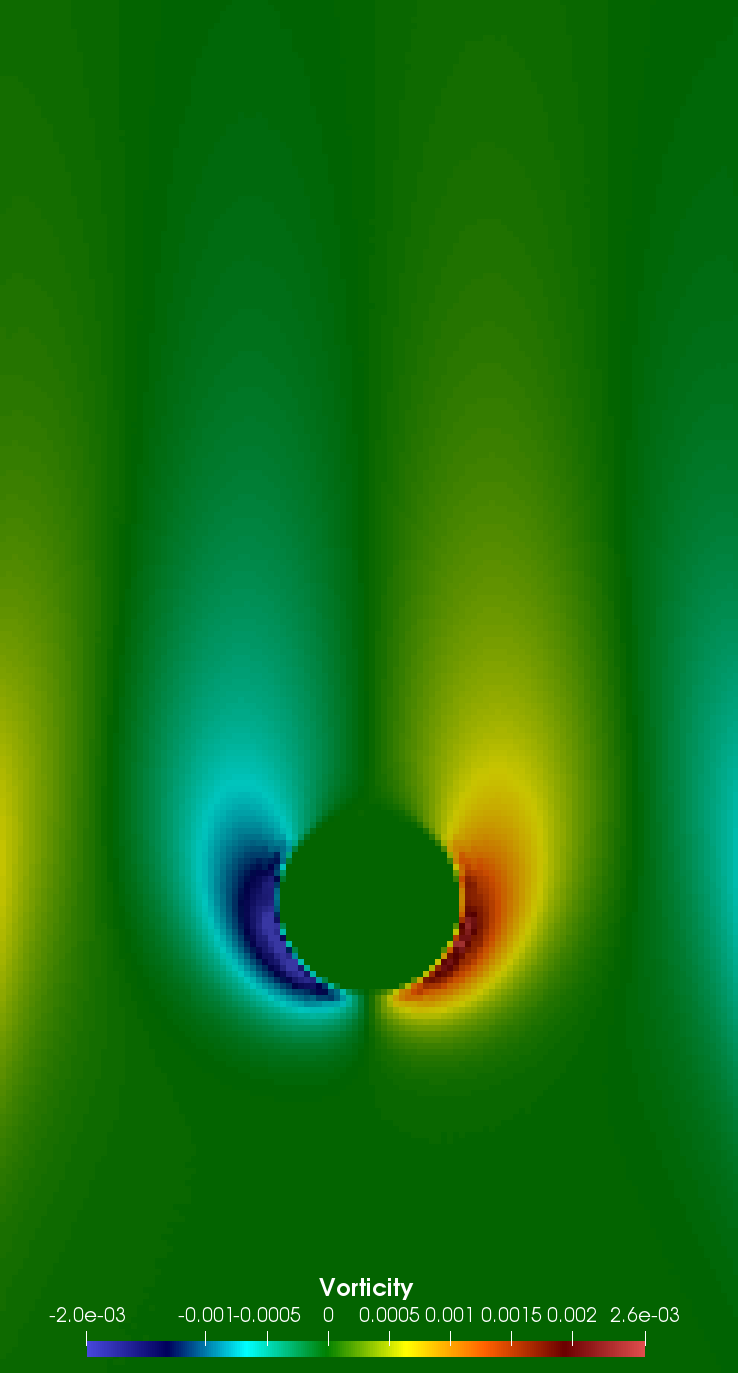}}
\caption{\label{vortnc} Fluid vorticity  at times t=0.4, 06, 1.0 and 3.0 seconds in lattice units}
\end{figure}

Figures \ref{vnc} and \ref{vortnc} show the fluid velocity and the vorticity field around the particle at four different times. The dynamics of the flow field and the particle can be analyzed using the velocity magnitude and the vorticity. The particle goes first to the right and rotates in a positive direction. Next a brief oscillation occurs around the central line of the channel and finally the particle stays in the middle of the channel with a steady velocity.

This example shows that the VP-LBM method is able to predict a complex trajectory for a real case of fluid structure interaction at a very low Reynolds number. 
\subsection{Vortex Induced Vibration of a cylinder}

%The last example is Vortex Induced Vibration (VIV) of a cylinder. 

Let us consider the case depicted in figure \ref{fig:cylindre}. In this paragraph the displacement of the cylinder is not imposed, but driven by the fluid forces. The displacement is let free according to the $\mathbf{y}$ axis. Initially, the cylinder is at rest, the force exerted by the spring on the cylinder is equal to $0$. Due to the vortex shedding, the fluid applies a force according to the vertical axis and the cylinder begins to oscillate. The rigid displacement of the cylinder is solved using equation (\ref{xstru}) projected onto the $\mathbf{y}$ axis, with $c=0$ and $\boldsymbol{\mathcal{F}_{ext}}=0$ : 
\begin{equation}
  m\ddot{y}_G+ k \left(y_G-y_0  \right)=\mathcal{F}_y
  \label{eq:solide}
\end{equation}
where $y_G$ is the position of the center of gravity of the cylinder according to the $\mathbf{y}$ axis, $y_0$ is the position at rest, and $\mathcal{F}_y$ denotes the fluid forces acting on the cylinder in the vertical direction. Equation (\ref{thetastru}) is not used. 

To simplify the analysis, the non dimensional form of equation (\ref{eq:solide}) is considered :
\begin{equation}
  m^{\ast} \ddot{y}^{\ast} + k^{\ast}\left( y^{\ast}-y^{\ast}_0 \right) = C_L
\label{eq:solideadim}
\end{equation}
with the non dimensional numbers :
$$
m^{\ast}=\dfrac{m}{0.5 \rho D^2} \hspace{0.5cm} k^{\ast}=\dfrac{k}{0.5 \rho U^2_{0}} \hspace{0.5cm} y^{\ast}=\dfrac{y_G}{D}  \hspace{0.5cm} C_L=\displaystyle \dfrac{\mathcal{F}_y}{\dfrac{1}{2} \rho U^2_{0}D}  \hspace{0.5cm}t^\ast=t\dfrac{U_{0}}{D}
$$

%The adimensional pulsation $\omega^{\ast}$ is introduced :
%\begin{equation}
% \omega^{\ast} =  \dfrac{k^{\ast}}{m^{\ast}}=\dfrac{k}{m}\dfrac{L^2}{U^2_{\infty}}
%  \label{eq:adimpuls}
%\end{equation}
To scale the results, the effective stiffness introduced by Shiels et al. \cite{Shiels20013} is used :
\begin{equation}
 k^\ast_{\textrm{eff}}=k^\ast -m^\ast {\omega^\ast} ^2 
\end{equation}
where $\omega^{\ast}$ is computed from the analysis of the cylinder displacement. The effective stiffness enables to use only one plot to represent the results, because it combines the reduced mass $m^\ast$ and the reduced stiffness $k^\ast$. For a given Reynolds number, $k^\ast_{\textrm{eff}}$ completely determines the system. 

The computational parameters are  identical to those related to the first test case (section \ref{sec:app1})
$$\tau = 0.56 \hspace{0.5cm} U_0=0.048780  \;\textrm{l.u.}  \hspace{0.5cm}  L_1 =1230\;\textrm{l.u.}  \hspace{0.5cm} L_2= 410 \;\textrm{l.u.}    \hspace{0.5cm} D=41 \;\textrm{l.u.} $$ 
 $29$ values of the stiffness parameter $k^\ast$ were considered, from $1$ to $30$. For each $k^\ast$, the choice of $\omega^{\ast}$ was based on the the analysis of the drag coefficient, and with these values, $k^\ast_{\textrm{eff}}$ was computed. 
\begin{figure}[!htbp]
  \centering
  %\subfigure[Initial grid for the VP and UIBB method \label{viva}]{\includegraphics[width=6cm]{figure/comparaisonR0.png}}
  %\subfigure[one grid reffinement \label{vivb} ]{\includegraphics[width=6cm]{figure/comparaisonR1.png}}
  \includegraphics[width=8cm]{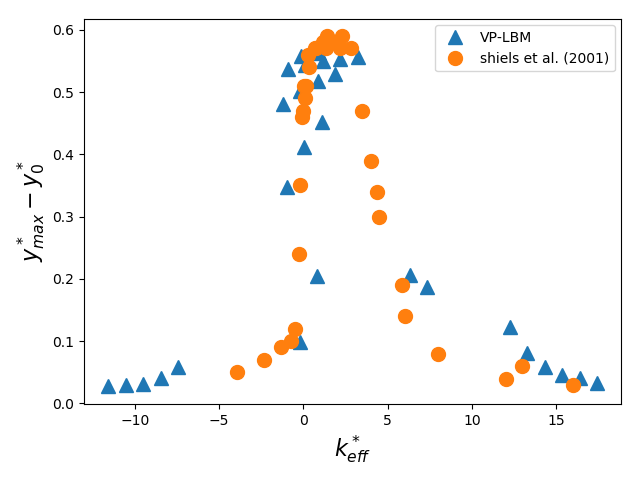}
  \caption{Results obtained  using the VP-LBM  and compared with Shiels et al.'s results \cite{Shiels20013}}
  \label{fig:viv}
\end{figure}

\begin{figure}[!htbp]
\center
\subfigure[$k^\ast=9 \; k^\ast_{\textrm{eff}}=0.058 $]{\includegraphics[width=4.3cm]{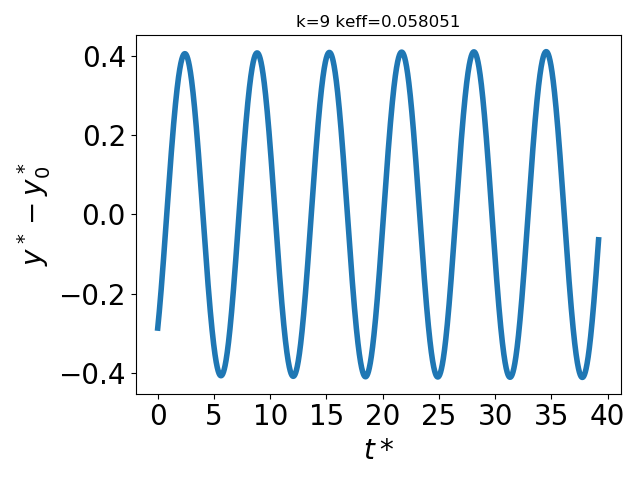}}
\subfigure[$k^\ast=19 \; k^\ast_{\textrm{eff}}=3.20 $]{\includegraphics[width=4.3cm]{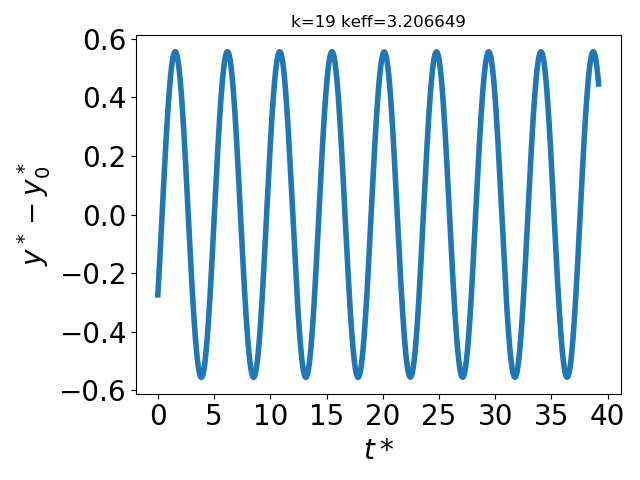}}
\subfigure[\label{poskb} $k^\ast=28 \; k^\ast_{\textrm{eff}}=16.37 $]{\includegraphics[width=4.3cm]{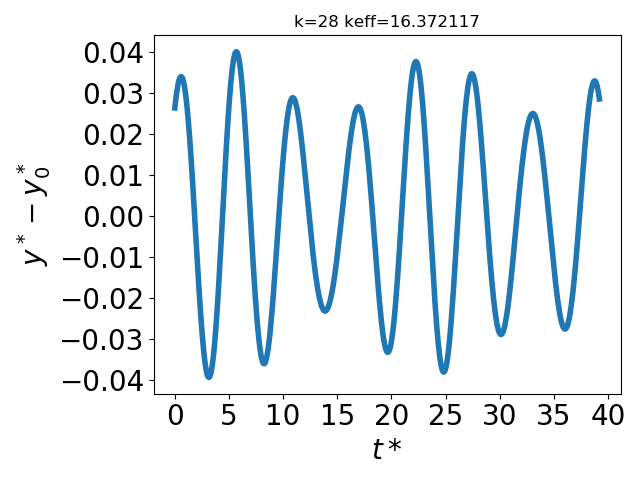}}
\caption{\label{posviv} Time-dependent value of the non dimensional amplitude $y^\ast - y^\ast_0$ for 3 values of $k^\ast$}
\end{figure}

\begin{figure}[!htbp]
\center
\subfigure[$k^\ast=9 \; k^\ast_{\textrm{eff}}=0.058 $]{\includegraphics[width=4.3cm]{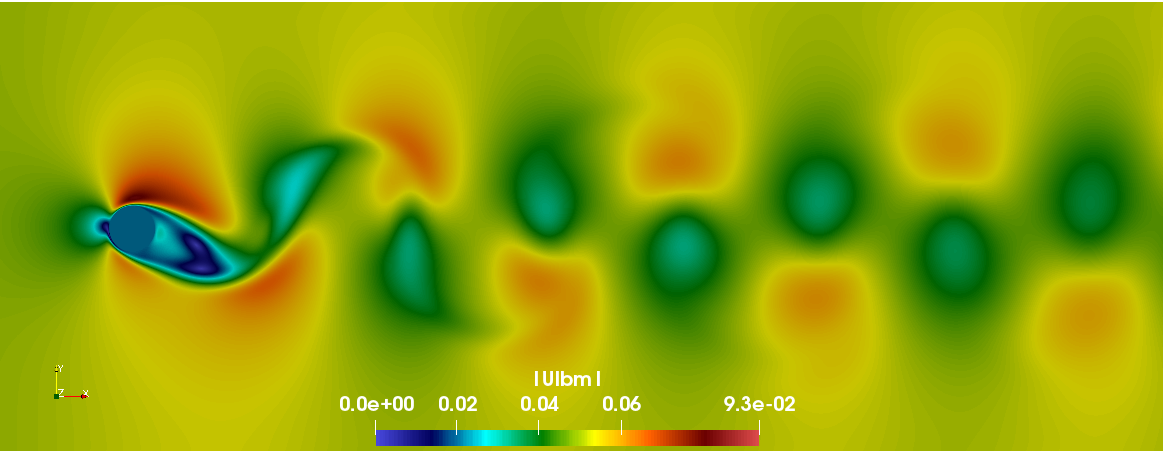}}
\subfigure[\label{vk19}$k^\ast=19 \; k^\ast_{\textrm{eff}}=3.20 $]{\includegraphics[width=4.3cm]{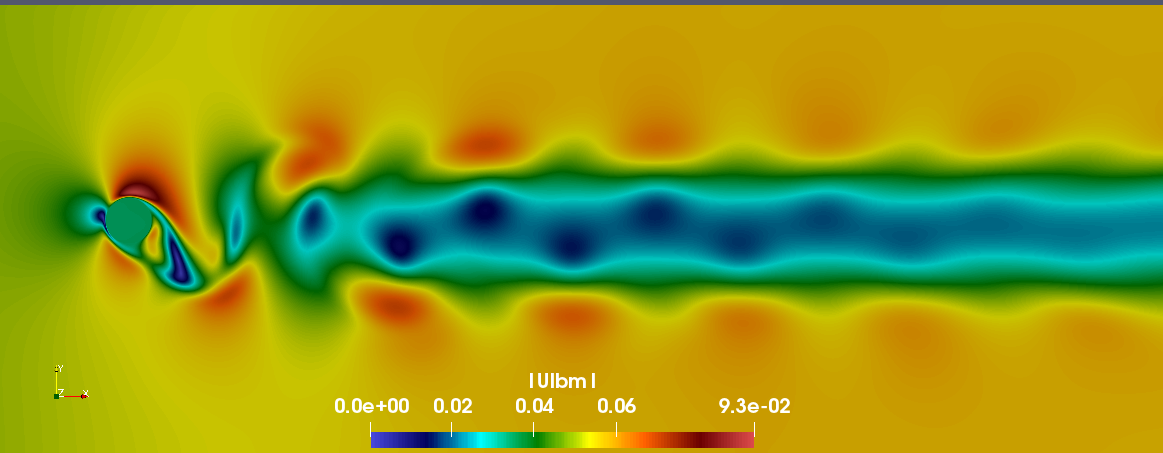}}
\subfigure[$k^\ast=28 \; k^\ast_{\textrm{eff}}=16.37 $]{\includegraphics[width=4.3cm]{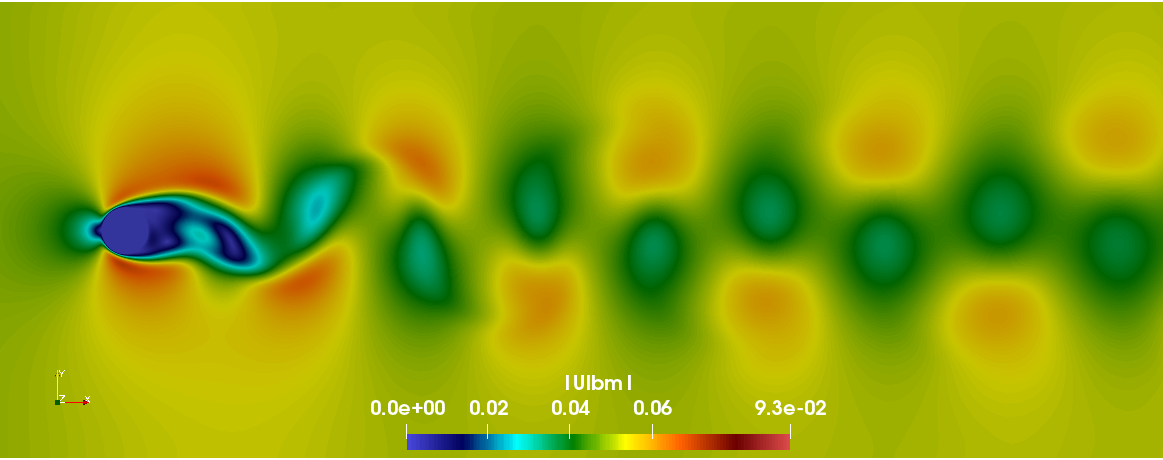}}
\caption{\label{velovivi} Field of velocity magnitude  in lattice units for 3 values of $k^\ast$}
\end{figure}

\begin{figure}[!htbp]
\center
\subfigure[$k^\ast=9 \; k^\ast_{\textrm{eff}}=0.058 $]{\includegraphics[width=4.3cm]{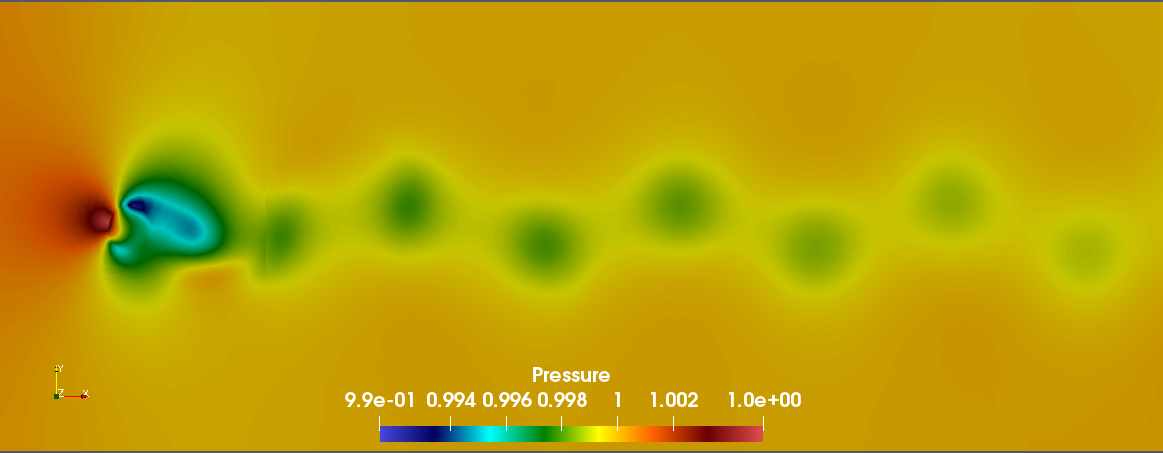}}
\subfigure[$k^\ast=19 \; k^\ast_{\textrm{eff}}=3.20 $]{\includegraphics[width=4.3cm]{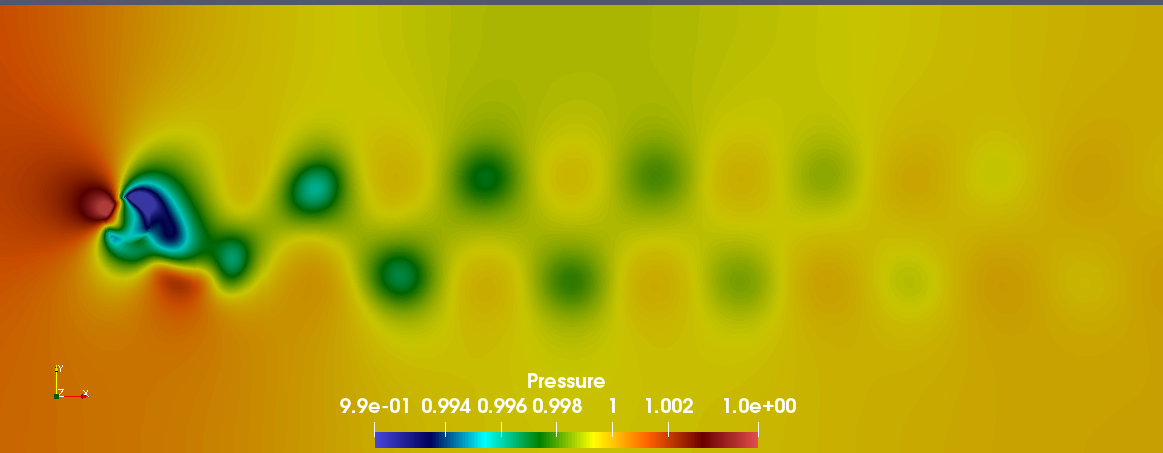}}
\subfigure[$k^\ast=28 \; k^\ast_{\textrm{eff}}=16.37 $]{\includegraphics[width=4.3cm]{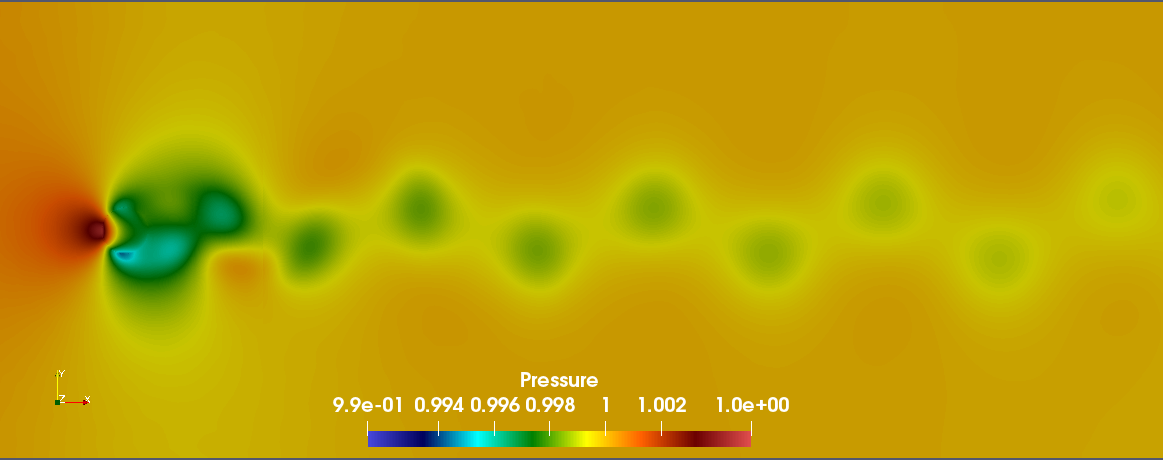}}
\caption{\label{pvivi} Pressure field in lattice units for 3 values of $k^\ast$}
\end{figure}

\begin{figure}[!htbp]
\center
\subfigure[$k^\ast=9 \; k^\ast_{\textrm{eff}}=0.058 $]{\includegraphics[width=4.3cm]{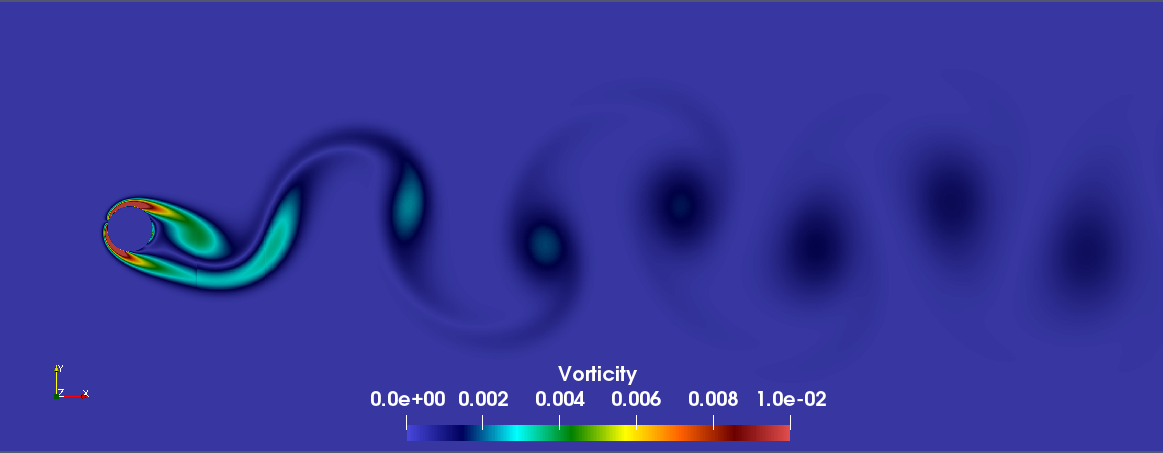}}
\subfigure[$k^\ast=19 \; k^\ast_{\textrm{eff}}=3.20 $]{\includegraphics[width=4.3cm]{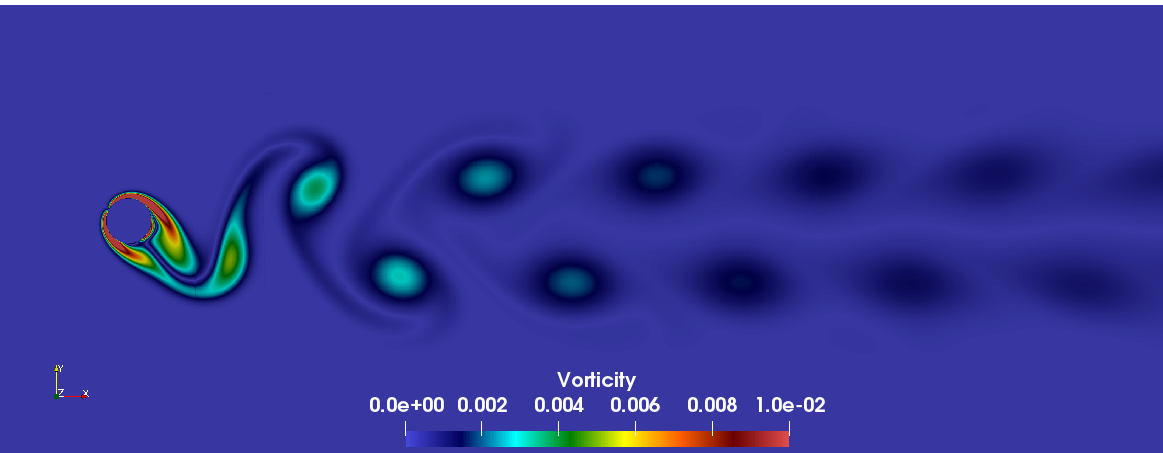}}
\subfigure[$k^\ast=28 \; k^\ast_{\textrm{eff}}=16.37 $]{\includegraphics[width=4.3cm]{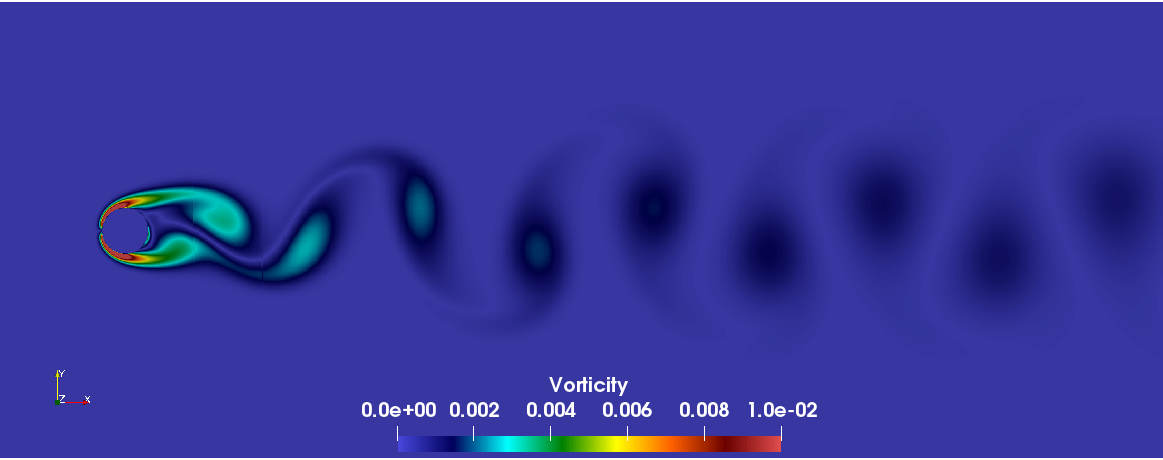}}
\caption{\label{vortvivi} Vorticity field in lattice units for 3 values of $k^\ast$}
\end{figure}

In figure \ref{fig:viv} the maximum of the non dimensional amplitude $y^\ast_{max}- y^\ast_0$ versus $k^\ast_{\textrm{eff}}$, obtained obtained by the VP-LBM methods and compared with data from \cite{Shiels20013} is plotted. It can be seen that the proposed method is able to reproduce the lock-in phenomena when $k^\ast$ is close to $m {\omega^\ast}^2$, and when important displacement of the cylinder occurs (more than $50 \%$ of the diameter). Figure \ref{posviv} depicts the temporal evolution of the non-dimensional amplitude for $3$ values of $k^\ast$. The velocity, pressure and vorticity fields are presented in figures \ref{velovivi}-\ref{vortvivi}. Different behaviors according to the value of $k^\ast$ were obtained. For high values of $k^\ast$ the displacement of the cylinder is small, and the fluid flow shows similar features as seen for a fixed cylinder. However, the behavior of the solid is more chaotic, several frequencies can be noticed when examining the displacement of the body (figure \ref{poskb}).
For $ 1 \leq k^\ast_{eff} \leq 5 $ the flow pattern changes drastically. The vortex shedding frequency increases and the velocity decreases in the wake of the cylinder (figure \ref{vk19}).


To conclude, this last application shows the capacity of the VP-LBM method to capture various physical behaviors induced by changes of the parameter $k^\ast$.

\section{Conclusion}
A combined approach coupling the Volume Penalization and the Lattice Boltzmann method for fluid structure interaction was proposed. The method consists in adding a force term, which is similar to Darcy's law, into the Boltzmann equation. The advantage of this method is that no explicit computation of the fluid structure interface is needed, the use of a characteristic function is sufficient. In addition, the fluid forces exerted on the structure are computed using the classical momentum exchange method. The method was implemented on a GPU architecture device and tested on three cases. The first case which deals with the imposed displacement of a cylinder in a transverse fluid flow at a Reynolds of $100$, validates the capacity of the method to compute drag and lift forces. The second application focuses on the sedimentation of a particle at a very low Reynolds number in a channel. The proposed method succeeds to capture the complex trajectory of the particle, which is composed of translational and rotational components. In the last application, the vortex induced vibration of a cylinder in a transverse fluid flow is considered. The capacity of the method to predict the physics of the fluid flow  and the structure behavior for different values of the stiffness parameter of the cylinder is tested. Results are in a good agreement with those of literature. All these applications validate the VP-LBM as an efficient tool to model FSI in case of rigid bodies.


\section{Bibliography}
\bibliographystyle{elsarticle-num} 
\bibliography{biblbmvp}

\begin{thebibliography}{10}
\expandafter\ifx\csname url\endcsname\relax
  \def\url#1{\texttt{#1}}\fi
\expandafter\ifx\csname urlprefix\endcsname\relax\def\urlprefix{URL }\fi
\expandafter\ifx\csname href\endcsname\relax
  \def\href#1#2{#2} \def\path#1{#1}\fi

\bibitem{Benzi1992145}
R.~Benzi, S.~Succi, M.~Vergassola, The lattice {Boltzmann} equation: theory and
  applications, Physics Reports 222~(3) (1992) 145--197.
\newblock \href {http://dx.doi.org/10.1016/0370-1573(92)90090-M}
  {\path{doi:10.1016/0370-1573(92)90090-M}}.

\bibitem{Fan2004297}
Z.~Fan, F.~Qiu, A.~Kaufman, S.~Yoakum-Stover, {GPU} cluster for high
  performance computing, IEEE/ACM SC2004 Conference, Proceedings (2004)
  297--308.

\bibitem{krugerbook}
T.~Kruger, H.~Kusumaatmaja, A.~Kuzmin, O.~Shardt, G.~Silva, E.~M. Viggen, The
  {Lattice} {Boltzmann} {Method} - Principles and Practice, Graduate Texts in
  Physics, Springer International Publishing, 2017.
\newblock \href {http://dx.doi.org/10.1007/978-3-319-44649-3}
  {\path{doi:10.1007/978-3-319-44649-3}}.

\bibitem{Ladd20011191}
A.~Ladd, R.~Verberg, Lattice-{Boltzmann} simulations of particle-fluid
  suspensions, Journal of Statistical Physics 104~(5-6) (2001) 1191--1251.
\newblock \href {http://dx.doi.org/10.1023/A:1010414013942}
  {\path{doi:10.1023/A:1010414013942}}.

\bibitem{Yu20032003}
D.~Yu, R.~Mei, W.~Shyy, A unified boundary treatment in lattice {Boltzmann}
  method. 41st aerospace sciences meeting and exhibit, vol. 1, AIAA (2003)
  2003--2953\href {http://dx.doi.org/10.2514/6.2003-953}
  {\path{doi:10.2514/6.2003-953}}.

\bibitem{Bouzidi20013452}
M.~Bouzidi, M.~Firdaouss, P.~Lallemand, Momentum transfer of a
  {Boltzmann-lattice} fluid with boundaries, Physics of Fluids 13~(11) (2001)
  3452--3459.
\newblock \href {http://dx.doi.org/10.1063/1.1399290}
  {\path{doi:10.1063/1.1399290}}.

\bibitem{Noble19981189}
D.~Noble, J.~Torczynski, A lattice-{Boltzmann} method for partially saturated
  computational cells, International Journal of Modern Physics C 9~(8) (1998)
  1189--1201.
\newblock \href {http://dx.doi.org/10.1142/S0129183198001084}
  {\path{doi:10.1142/S0129183198001084}}.

\bibitem{Feng2004602}
Z.-G. Feng, E.~Michaelides, The immersed boundary-lattice {Boltzmann} method
  for solving fluid-particles interaction problems, Journal of Computational
  Physics 195~(2) (2004) 602--628.
\newblock \href {http://dx.doi.org/10.1016/j.jcp.2003.10.013}
  {\path{doi:10.1016/j.jcp.2003.10.013}}.

\bibitem{Dupuis20084486}
A.~Dupuis, P.~Chatelain, P.~Koumoutsakos, An immersed
  boundary-lattice-{Boltzmann} method for the simulation of the flow past an
  impulsively started cylinder, Journal of Computational Physics 227~(9) (2008)
  4486--4498.
\newblock \href {http://dx.doi.org/10.1016/j.jcp.2008.01.009}
  {\path{doi:10.1016/j.jcp.2008.01.009}}.

\bibitem{Wang2015440}
Y.~Wang, C.~Shu, C.~Teo, J.~Wu, An immersed boundary-lattice {Boltzmann} flux
  solver and its applications to fluid structure interaction problems, Journal
  of Fluids and Structures 54 (2015) 440 -- 465.
\newblock \href {http://dx.doi.org/10.1016/j.jfluidstructs.2014.12.003}
  {\path{doi:10.1016/j.jfluidstructs.2014.12.003}}.

\bibitem{Angot1999497}
P.~Angot, C.-H. Bruneau, P.~Fabrie, A penalization method to take into account
  obstacles in incompressible viscous flows, Numerische Mathematik 81~(4)
  (1999) 497--520.
\newblock \href {http://dx.doi.org/10.1007/s002110050401}
  {\path{doi:10.1007/s002110050401}}.

\bibitem{Benamour2015299}
M.~Benamour, E.~Liberge, C.~B\'eghein, Lattice {Boltzmann} method for fluid
  flow around bodies using volume penalization, International Journal of
  Multiphysics 9~(3) (2015) 299--315.
\newblock \href {http://dx.doi.org/10.1260/1750-9548.9.3.299}
  {\path{doi:10.1260/1750-9548.9.3.299}}.

\bibitem{Benamour2017481}
M.~Benamour, E.~Liberge, C.~B\'eghein, A new approach using lattice {Boltzmann}
  method to simulate fluid structure interaction, Energy Procedia 139 (2017)
  481--486.
\newblock \href {http://dx.doi.org/10.1016/j.egypro.2017.11.241}
  {\path{doi:10.1016/j.egypro.2017.11.241}}.

\bibitem{Kadoch20124365}
B.~Kadoch, D.~Kolomenskiy, P.~Angot, K.~Schneider, A volume penalization method
  for incompressible flows and scalar advection-diffusion with moving
  obstacles, Journal of Computational Physics 231~(12) (2012) 4365--4383.
\newblock \href {http://dx.doi.org/10.1016/j.jcp.2012.01.036}
  {\path{doi:10.1016/j.jcp.2012.01.036}}.

\bibitem{Bhatnagar1954511}
P.~Bhatnagar, E.~Gross, M.~Krook, A model for collision processes in gases. {I.
  Small} amplitude processes in charged and neutral one-component systems,
  Physical Review 94~(3) (1954) 511--525.
\newblock \href {http://dx.doi.org/10.1103/PhysRev.94.511}
  {\path{doi:10.1103/PhysRev.94.511}}.

\bibitem{dh92}
D.~d'Humi\`ere, Rarefied Gas Dynamics: Theory and Simulations, Progress in
  Astronautics and Aeronautics, 1992, Ch. Generalized Lattice-Boltzmann
  Equations, pp. 450--458.
\newblock \href {http://dx.doi.org/10.2514/5.9781600866319.0450.0458}
  {\path{doi:10.2514/5.9781600866319.0450.0458}}.

\bibitem{Lu2012}
J.~Lu, H.~Han, B.~Shi, Z.~Guo, Immersed boundary lattice {Boltzmann} model
  based on multiple relaxation times, Physical Review E - Statistical,
  Nonlinear, and Soft Matter Physics 85~(1).
\newblock \href {http://dx.doi.org/10.1103/PhysRevE.85.016711}
  {\path{doi:10.1103/PhysRevE.85.016711}}.

\bibitem{Yu2003329}
D.~Yu, R.~Mei, L.-S. Luo, W.~Shyy, Viscous flow computations with the method of
  lattice {Boltzmann} equation, Progress in Aerospace Sciences 39~(5) (2003)
  329--367.
\newblock \href {http://dx.doi.org/10.1016/S0376-0421(03)00003-4}
  {\path{doi:10.1016/S0376-0421(03)00003-4}}.

\bibitem{guo_discrete_2002}
Z.~Guo, C.~Zheng, B.~Shi, Discrete lattice effects on the forcing term in the
  lattice {Boltzmann} method, Physical Review E 65~(4) (2002) 046308.
\newblock \href {http://dx.doi.org/10.1103/PhysRevE.65.046308}
  {\path{doi:10.1103/PhysRevE.65.046308}}.

\bibitem{WEN2014161}
B.~Wen, C.~Zhang, Y.~Tu, C.~Wang, H.~Fang, Galilean invariant fluid–solid
  interfacial dynamics in lattice {Boltzmann} simulations, Journal of
  Computational Physics 266 (2014) 161 -- 170.
\newblock \href {http://dx.doi.org/10.1016/j.jcp.2014.02.018}
  {\path{doi:10.1016/j.jcp.2014.02.018}}.

\bibitem{PhysRevE.92.063302}
J.~P. Giovacchini, O.~E. Ortiz, Flow force and torque on submerged bodies in
  lattice-{Boltzmann} methods via momentum exchange, Phys. Rev. E 92 (2015)
  063302.
\newblock \href {http://dx.doi.org/10.1103/PhysRevE.92.063302}
  {\path{doi:10.1103/PhysRevE.92.063302}}.

\bibitem{codesaturne}
F.~Archambeau, N.~Méchitoua, S.~M., Code\_saturne : a finite volume code for
  the computation of turbulent incompressible flows, International Journal on
  Finite Volumes 1.

\bibitem{Yang2013160}
Z.~Yang, Lattice {Boltzmann} outflow treatments: Convective conditions and
  others, Computers and Mathematics with Applications 65~(2) (2013) 160--171.
\newblock \href {http://dx.doi.org/10.1016/j.camwa.2012.11.012}
  {\path{doi:10.1016/j.camwa.2012.11.012}}.

\bibitem{Wang20131151}
L.~Wang, Z.~Guo, B.~Shi, C.~Zheng, Evaluation of three lattice {Boltzmann}
  models for particulate flows, Communications in Computational Physics 13~(4)
  (2013) 1151--1171.
\newblock \href {http://dx.doi.org/10.4208/cicp.160911.200412a}
  {\path{doi:10.4208/cicp.160911.200412a}}.

\bibitem{Tao20161}
S.~Tao, J.~Hu, Z.~Guo, An investigation on momentum exchange methods and
  refilling algorithms for lattice {Boltzmann} simulation of particulate flows,
  Computers and Fluids 133 (2016) 1--14.
\newblock \href {http://dx.doi.org/10.1016/j.compfluid.2016.04.009}
  {\path{doi:10.1016/j.compfluid.2016.04.009}}.

\bibitem{happel}
J.~Happel, H.~Brenner, Low {Reynolds} number hydrodynamics, Springer
  Netherlands, 1983.
\newblock \href {http://dx.doi.org/10.1007/978-94-009-8352-6}
  {\path{doi:10.1007/978-94-009-8352-6}}.

\bibitem{Dorschner2015340}
B.~Dorschner, S.~Chikatamarla, F.~BÃ¶sch, I.~Karlin, Grad's approximation for
  moving and stationary walls in entropic lattice {Boltzmann} simulations,
  Journal of Computational Physics 295 (2015) 340--354.
\newblock \href {http://dx.doi.org/10.1016/j.jcp.2015.04.017}
  {\path{doi:10.1016/j.jcp.2015.04.017}}.

\bibitem{Shiels20013}
D.~Shiels, A.~Leonard, A.~Roshko, Flow-induced vibration of a circular cylinder
  at limiting structural parameters, Journal of Fluids and Structures 15~(1)
  (2001) 3--21.
\newblock \href {http://dx.doi.org/10.1006/jfls.2000.0330}
  {\path{doi:10.1006/jfls.2000.0330}}.

\end{thebibliography}


%\begin{thebibliography}{00}



%\end{thebibliography}

\end{document}